\documentstyle[epsf,aps]{revtex}
\gdef\Feynmanlength{\setlength{\unitlength}{0.01pt}}  % Say \Feynmanlength

\global\newcount\LINETYPE
\global\newcount\LINEDIRECTION
\global\newcount\LINECONFIGURATION
\newcommand{\LTYPE}{\LINETYPE}
\newcommand{\LDIR}{\LINEDIRECTION}

\global\LINETYPE=1  \global\LINEDIRECTION=0  \global\LINECONFIGURATION=0
\global\newcount\fermion    \fermion=1
\global\newcount\scalar     \scalar=2
\global\newcount\photon     \photon=3
\global\newcount\gluon      \gluon=4
\global\newcount\SPECIAL    \SPECIAL=5
    \gdef\E{2}   
\gdef\S{4}       
\global\newcount\REG            \global\REG=0
\global\newcount\FLIPPED        \global\FLIPPED=1
\global\newcount\CURLY          \global\CURLY=2
\global\newcount\FLIPPEDCURLY   \global\FLIPPEDCURLY=3
\global\newcount\FLAT           \global\FLAT=4
\global\newcount\FLIPPEDFLAT    \global\FLIPPEDFLAT=5
\global\newcount\CENTRAL        \global\CENTRAL=6
\global\newcount\FLIPPEDCENTRAL \global\FLIPPEDCENTRAL=7
             
\global\newcount\SQUASHEDGLUON  \global\SQUASHEDGLUON=8

\newcount\adjx \adjx=0
\newcount\adjy \adjy=0
\global\newdimen\BIGPHOTONS     \BIGPHOTONS=0pt  %  DEFAULT:  10 & 11-PT
\global\newdimen\THICKPHOTONS     \THICKPHOTONS=0pt  %  FOR E-W PHOTONS
\global\newdimen\THICKPHOTONSWITCH    \THICKPHOTONSWITCH=0pt
\gdef\THICKPHOTONTEST{
\THICKPHOTONSWITCH=0pt
\ifdim\THICKPHOTONS=0pt \relax
  \else \ifnum\LTYPE=3
           \ifnum\LDIR=2 \THICKPHOTONSWITCH=1pt \fi % THICK \E PHOTON
           \ifnum\LDIR=6 \THICKPHOTONSWITCH=1pt \fi % THICK \W PHOTON
        \fi
\fi
}  % end of THICKPHOTONTEST
\gdef\THICKLINES{\thicklines  \THICKPHOTONS=1pt}

\global\newcount\phantomswitch   \global\phantomswitch=0
\global\newcount\stemlength   \global\stemlength=275   % Default STEM length.
\global\newcount\absstemlength        % A copy of STEM length.
\global\newcount\stemlengthx          % FOR STEMS on particle lines
\global\newcount\stemlengthy          % FOR STEMS on particle lines
\newdimen\FRONTSTEM  \FRONTSTEM=0pt   % FOR STEMS
\newdimen\BACKSTEM   \BACKSTEM=0pt    % FOR STEMS
\newdimen\EITHERSTEM \EITHERSTEM=0pt  % FOR STEMS
\global\newcount\arrowlength                % FOR ARROWS
\global\newdimen\ATTIP   \global\ATTIP=0pt  % FOR ARROWS
\global\newdimen\ATBASE  \global\ATBASE=1pt % FOR ARROWS
\global\newcount\unitboxnumber  % SHOWS THE NUMBER OF `UNIT BOXES' IN LINE
\global\newcount\unitboxnumberpo  % One more than \unitboxnumber (in
\global\newcount\particlelengthx  % THE X-LENGTH OF THE PARTICLE LINE
\gdef\plengthx{\particlelengthx}
\global\newcount\particlelengthy  % THE Y-LENGTH OF THE PARTICLE LINE
\gdef\plengthy{\particlelengthy}
\global\newcount\boxlengthx  % THE X-LENGTH OF THE BOX:  abs(plengthx) usually
\global\newcount\boxlengthy  % THE y-LENGTH OF THE box:  abs(plengthy) usually
\global\newcount\particleadjustx  % Replaces \gluonadjustx, \scalaradjustx etc.
\global\newcount\particleadjusty  % Replaces \gluonadjusty, \scalaradjusty etc.
\global\newcount\particlelength   % The LENGTH of a particle line BOX (x)
\global\newcount\particlefrontx
\gdef\pfrontx{\particlefrontx}
\global\newcount\PFRONTx
\global\newcount\particlefronty
\gdef\pfronty{\particlefronty}
\global\newcount\PFRONTy
\global\newcount\particlebackx
\gdef\pbackx{\particlebackx}
\global\newcount\particlebacky
\gdef\pbacky{\particlebacky}
\global\newcount\particlemidx
\gdef\pmidx{\particlemidx}
\global\newcount\particlemidy
\gdef\pmidy{\particlemidy}
\global\newcount\seglength  \global\newcount\gaplength
\global\gaplength=850  %default
\global\seglength=1416  % Length of each seg not including `ends' for
\global\newcount\Xone    \global\newcount\Yone    % user co-ords (\Xone,\Yone)
\global\newcount\Xtwo    \global\newcount\Ytwo    % user co-ords (\Xtwo,\Ytwo)
\global\newcount\Xthree  \global\newcount\Ythree  % user's (\Xthree,\Ythree)
\global\newcount\Xfour   \global\newcount\Yfour   % user co-ords
\global\newcount\Xfive   \global\newcount\Yfive   % user co-ords
\global\newcount\Xsix    \global\newcount\Ysix    % user co-ords (\Xsix,\Ysix)
\global\newcount\Xseven  \global\newcount\Yseven  % user's (\Xseven,\Yseven)
\global\newcount\Xeight  \global\newcount\Yeight  % user's (\Xeight,\Yeight)
\newsavebox{\lastline}  %  Default name for an unnamed particle line.
\global\newcount\numlineparts   % Num of pieces each unitbox of the line needs
\global\newcount\upperlineadjx  \upperlineadjx=0  %Default
\global\newcount\upperlineadjy  \upperlineadjy=0  %Default
\global\newcount\lowerlineadjx  \lowerlineadjx=0  %Default
\global\newcount\lowerlineadjy  \lowerlineadjy=0  %Default
\global\newcount\thirdlineadjx  \thirdlineadjx=0  %Default
\global\newcount\thirdlineadjy  \thirdlineadjy=0  %Default
\global\newcount\fourthlineadjx \fourthlineadjx=0  %Default
\global\newcount\fourthlineadjy \fourthlineadjy=0  %Default
\global\newcount\unitboxwidth   \unitboxwidth=1000%Default
\global\newcount\unitboxheight  \unitboxheight=0  %Default
\global\newcount\numupperunits  \numupperunits=8  %Default
\global\newcount\numlowerunits  \numlowerunits=8  %Default
\global\newcount\numthirdunits  \numthirdunits=8  %Default
\global\newcount\numfourthunits \numfourthunits=8  %Default
\global\newcount\fermioncount   \global\fermioncount=0
\global\newcount\scalarcount    \global\scalarcount=0
\global\newcount\photoncount    \global\photoncount=0
\global\newcount\gluoncount     \global\gluoncount=0
\global\newcount\SPECIALcount   \global\SPECIALcount=0
\global\newcount\vertexcount    \global\vertexcount=-1
\global\newcount\XDIR
\global\newcount\YDIR
\gdef\SETDIR{  % SETS THE DIRECTIONS
\ifcase\LDIR
     \global\XDIR=0  \global\YDIR=1   %\N  case.
\or  \global\XDIR=1  \global\YDIR=1   %\NE case.
\or  \global\XDIR=1  \global\YDIR=0   %\E  case.
\or  \global\XDIR=1  \global\YDIR=-1  %\SE case.
\or  \global\XDIR=0  \global\YDIR=-1  %\S  case.
\or  \global\XDIR=-1 \global\YDIR=-1  %\SW case.
\or  \global\XDIR=-1 \global\YDIR=0   %\W  case.
\or  \global\XDIR=-1 \global\YDIR=1   %\NW case.
\else\DIRECTERROR
\fi}  % END OF \SETDIR
\gdef\moduloeight#1{
\ifnum#1>7 \global\advance #1 by -8
\relax
\moduloeight#1
\relax
\else \relax
\fi}
\gdef\multroothalf#1{\global\multiply #1 by 7071 \global\divide #1 by 10000}
\gdef\negate#1{\global\multiply #1 by -1}

\gdef\slanttest(#1,#2){
\ifodd\LDIR
\multiply #1 by 7071  \divide #1 by 10000
\multiply #2 by 7071  \divide #2 by 10000
\fi
}
\gdef\gslanttest(#1,#2){
\ifodd\LDIR
\multroothalf#1
\multroothalf#2
\fi
}
\gdef\setplength{ % calcs length of particle line
\global\particlelengthx=\unitboxwidth
\global\particlelengthy=\unitboxheight
\global\multiply \particlelengthx by \unitboxnumber
\global\multiply \particlelengthy by \unitboxnumber
\global\advance \particlelengthx by \particleadjustx
\global\advance \particlelengthy by \particleadjusty
}
\gdef\boxlengthdefault{  % DEFAULT FOR BOX SIZES IN \drawas
\global\boxlengthx=\plengthx
\global\boxlengthy=\plengthy
\ifnum\plengthx<0 \global\multiply\boxlengthx by -1 \fi
\ifnum\plengthy<0 \global\multiply\boxlengthy by -1 \fi
}
\gdef\rearcoords{  %  CALCULATES THE CO-ORDINATES OF THE BACK OF PARTICLE LINE
\global\particlebacky=\particlefronty
\global\particlebackx=\particlefrontx
\global\advance \particlebackx by \particlelengthx
\global\advance \particlebacky by \particlelengthy
}
\gdef\midcoords{  %  CALCULATES THE CO-ORDINATES OF THE MID OF PARTICLE LINE
\global\particlemidy=\particlefronty
\global\particlemidx=\particlefrontx
\global\stemlengthx=\particlelengthx  % Convenient variables not being used
\global\stemlengthy=\particlelengthy
\global\divide\stemlengthx by 2
\global\divide\stemlengthy by 2
\global\advance \particlemidx by \stemlengthx
\global\advance \particlemidy by \stemlengthy
}
\gdef\setparticle{\setplength\rearcoords\midcoords\boxlengthdefault}  %sets
\gdef\setcoords(#1,#2,#3)(#4,#5,#6)[#7,#8]{
\global\upperlineadjx=#1
\global\lowerlineadjx=#2
\global\thirdlineadjx=#3
\global\upperlineadjy=#4
\global\lowerlineadjy=#5
\global\thirdlineadjy=#6
\global\unitboxwidth=#7
\global\unitboxheight=#8
}
\gdef\drawoldpic#1(#2,#3){  % DRAWS PRE-SAVED PICTURE
\global\particlefrontx=#2
\global\particlefronty=#3
\rearcoords
\midcoords
\put(#2,#3){\usebox{#1}}
}
\gdef\drawsavedline`#1' as #2[#3#4](#5,#6)[#7]{
\global\LINETYPE=#2
\global\LINEDIRECTION=#3
\global\LINECONFIGURATION=#4
\global\particlefrontx=#5
\global\particlefronty=#6
\global\unitboxnumber=#7
\selectcase
\rearcoords% moved from before selectcase.
\midcoords
\ifnum\phantomswitch=0 \drawas{#1}\fi
}

\gdef\drawas#1{
\global\savebox{#1}(\boxlengthx,\boxlengthy){
\setlength{\unitlength}{0.01pt}
\begin{picture}(\boxlengthx,\boxlengthy)
\multiput(\upperlineadjx,\upperlineadjy)(\unitboxwidth,\unitboxheight)
{\numupperunits}{\upperunitbox}
\ifnum\numlineparts > 1  %  If the line needs 2 parts per unit or more
\multiput(\lowerlineadjx,\lowerlineadjy)(\unitboxwidth,\unitboxheight)
{\numlowerunits}{\lowerunitbox}
\fi
\ifnum\numlineparts > 2  %  If the line needs 3 parts per unit or more
\multiput(\thirdlineadjx,\thirdlineadjy)(\unitboxwidth,\unitboxheight)
{\numthirdunits}{\thirdunitbox}
\fi
\ifnum\numlineparts > 3  %  If the line needs 4 parts per unit or more
\multiput(\fourthlineadjx,\fourthlineadjy)(\unitboxwidth,\unitboxheight)
{\numfourthunits}{\lowerunitbox}
\fi
\end{picture} }
\global\PFRONTx=\pfrontx  \global\PFRONTy=\pfronty   %save this value
\SETFRONTSTEM
\THICKPHOTONTEST
\ifdim\THICKPHOTONSWITCH=1pt\global\advance\PFRONTy by 20  \fi
\put(\PFRONTx,\PFRONTy) {\usebox{#1}}   %\pfrontX,Y=\particlefrontx,y
\ifdim\THICKPHOTONSWITCH=1pt
\global\advance\PFRONTy by -40
\put(\PFRONTx,\PFRONTy) {\usebox{#1}}   % The second \E or \W photon ->thicker
\global\advance \PFRONTy by 20  %re-adjust:  advanced by -20 in total above.
\fi  %End of \ifdim\THICKPHOTONSWITCH=1
\SETBACKSTEM
\seglength=1416   \gaplength=850   % Re-set \SCALR defaults.
}
\gdef\drawandsaveline`#1' as #2[#3#4](#5,#6)[#7]{
\global\newsavebox{#1}
\drawsavedline`#1' as #2[#3#4](#5,#6)[#7]
}

\gdef\drawline#1[#2#3](#4,#5)[#6]{   % Draw line but don't name it.
\drawsavedline`\lastline' as #1[#2#3](#4,#5)[#6]}

\gdef\TYPEERROR{\message{*** ERROR IN PARTICLE TYPE SELECTION ***}
\message{+++ Try with line type \fermion,\scalar,\photon,\gluon
(see manual) +++}\SETERR}
\gdef\DIRECTERROR{\SETERR\message{*** ERROR IN PARTICLE DIRECTION SELECTION
***}
\message{+++ Try again with direction N, NE, E, SE  etc. or see manual +++}}
\gdef\UNIMPERROR{\message{*** ERROR IN PARTICLE OPTIONS SELECTION ***}
\message{
+++ The requested options combination has not yet been implemented +++}\SETERR}
\gdef\SETERR{\gdef\upperunitbox{{\tiny Error}}  % PRINTS `error' in diagram.
\gdef\lowerunitbox{\relax}
\gdef\thirdunitbox{\relax}
}
\gdef\neglengthcheck{\ifnum\unitboxnumber < 1
\message{   *** ERROR:  PARTICLE OF NEGATIVE OR ZERO LENGTH REQUESTED. ***   }
\message{   ***         TAKING ABSOLUTE VALUE. ***   }\negate\unitboxnumber
\fi}
\gdef\selectcase{
\neglengthcheck   %  check for particles of negative length.
\SETDIR
\ifcase\LINETYPE
\TYPEERROR  % \LINETYPE=0 case.
\or \selectfermion  % \LINETYPE=1 case.
\or \selectscalar   % \LINETYPE=2 case.
\or \selectphoton   % \LINETYPE=3 case.
\or \selectgluon    % \LINETYPE=4 case.
\or \selectspecial  % \LINETYPE=5 case.
\else \TYPEERROR \fi  }
\gdef\selectfermion{
\ifnum\fermioncount=0 %                        FERMIONSETUP(7).TEX
\global\newcount\fermionlength  %  THE TOTAL FERMION LINE LENGTH.
\global\newcount\fermionlengthx
\global\newcount\fermionlengthy
\global\newcount\fermionfrontx  %}(x,y) co-ord of left of fermion
\global\newcount\fermionfronty  %}
\global\newcount\fermionbackx
\global\newcount\fermionbacky
\gdef\ALLfermion{  % READ IN FROM FEYNMAN \selectfermion
\global\fermionfrontx=\particlefrontx \global\fermionfronty=\particlefronty
\ifnum\unitboxnumber > 50000
\message{   *** WARNING *** Fermion of length
\the\unitboxnumber\space requested ***   }
\ifnum\unitboxnumber > 80000
\message{   *** Reducing fermion length to 30000 (max 80000) ***   }
\global\unitboxnumber=30000 \fi \fi  % end of length error
\global\fermionlength=\unitboxnumber % The TOTAL line length
\global\particleadjustx=0   \global\particleadjusty=0 %Default
\global\numlineparts = 1    \global\numupperunits=1
\global\upperlineadjx=-200  \global\upperlineadjy=0
\global\fermionlengthx=\fermionlength    \global\fermionlengthy=\fermionlength
\gslanttest(\fermionlengthx,\fermionlengthy)  % See FEYNMAN22.TEX (FOR \XDIR).
\global\multiply\fermionlengthx by \XDIR  %  In keeping with photons and
\global\multiply\fermionlengthy by \YDIR  %  In keeping with photons and
\global\unitboxheight=\fermionlengthy   \global\unitboxwidth=\fermionlengthx
\global\advance \fermionlengthx by \particleadjustx
\global\advance \fermionlengthy by \particleadjusty
\global\particlelengthx=\fermionlengthx
\global\particlelengthy=\fermionlengthy
\boxlengthdefault    \rearcoords    \midcoords
\global\fermionbackx=\particlebackx     \global\fermionbacky=\particlebacky
\ifcase\LINECONFIGURATION  %\REG case
\ifnum\XDIR=0
\gdef\upperunitbox{\line(\XDIR,\YDIR){\boxlengthy}} %\N or \S
\else
\gdef\upperunitbox{\line(\XDIR,\YDIR){\boxlengthx}}
\fi
\else \UNIMPERROR
\fi
}

 \fi
\global\advance\fermioncount by 1  % Counts number of fermions drawn.
\ALLfermion
}
\gdef\selectscalar{
\ifnum\scalarcount=0 %                   SCALARSETUP(9).TEX
\newcount\scalarlength
\newcount\scalarlengthx
\newcount\scalarlengthy
\newcount\scalarfrontx  %}(x,y) co-ord of left of scalar
\newcount\scalarfronty  %}
\newcount\scalarbackx
\newcount\scalarbacky
\gdef\ALLscalar{
\global\scalarfrontx=\particlefrontx   % READ IN FROM FEYNMAN \selectscalar
\global\scalarfronty=\particlefronty   % READ IN FROM FEYNMAN \selectscalar
\numlineparts = 1      \numupperunits=\unitboxnumber
\ifcase\LINECONFIGURATION
\global\upperlineadjx=-200     \global\upperlineadjy=0
\slanttest(\seglength,\gaplength)   %SEE FEYNMAN22.TEX.
\gdef\upperunitbox{\line(\XDIR,\YDIR){\seglength}}
\else \UNIMPERROR % etc.
\fi
\global\unitboxwidth=\seglength  \global\advance\unitboxwidth by \gaplength
\global\multiply \unitboxwidth by \XDIR
\global\unitboxheight=\seglength  \global\advance\unitboxheight by \gaplength
\global\multiply \unitboxheight by \YDIR
\global\particleadjustx=\gaplength \global\multiply\particleadjustx by \XDIR
\global\particleadjusty=\gaplength \global\multiply\particleadjusty by \YDIR
\negate\particleadjustx   \negate\particleadjusty   % SUBTRACT from linelength
\setparticle  %SCALAR8
\global\scalarlengthx=\particlelengthx  %SCALAR8
\global\scalarlengthy=\particlelengthy  %SCALAR8
\ifnum\boxlengthx > 50000
\message{   *** WARNING *** Scalar of length in excess of 50000cp
requested!}\fi
\ifnum\boxlengthy > 50000
\message{   *** WARNING *** Scalar of length in excess of 50000cp
requested!}\fi
\global\scalarbackx=\pbackx      \global\scalarbacky=\pbacky   %SCALAR8
}

 \fi
\global\advance\scalarcount by 1  % Counts number of scalars drawn.
\ALLscalar
}
\gdef\selectphoton{   % RECURSIVELY RE-DEFINED IN PHOTONSETUP(23+).TEX.
\ifnum\photoncount=0 \input PHOTONSETUP  \fi
\selectphoton
}
\gdef\selectgluon{   % RECURSIVELY RE-DEFINED IN GLUONSETUP(25+).TEX.
\ifnum\gluoncount=0 \input GLUONSETUP  \fi
\selectgluon
}
\gdef\selectspecial{\UNIMPERROR}
\gdef\checkvertex{ %immediately re-defines
\ifnum\vertexcount=-1   \input VERTEX  \fi}
\gdef\drawvertex#1[#2#3](#4,#5)[#6]{\checkvertex\drawvertex#1[#2#3](#4,#5)[#6]}
\gdef\vertexcap#1{\checkvertex\vertexcap#1}
\gdef\vertexcaps{\checkvertex\vertexcaps}
\gdef\vertexlink#1{\checkvertex\vertexlink#1}
\gdef\vertexlinks{\checkvertex\vertexlinks}
\gdef\stemvertex#1{\checkvertex\stemvertex#1}
\gdef\stemvertices{\checkvertex\stemvertices}
\gdef\flipvertex{\checkvertex\flipvertex}
\global\arrowlength=349  % Length of arrow
\gdef\drawarrow[#1#2](#3,#4){
\global\LDIR=#1
\SETDIR
\global\boxlengthx=#3  % Just a convenient variable name.  No significance.
\global\boxlengthy=#4  % The Arrow co-ordinates.
\ifdim#2=1pt  % CASE \ATBASE WHERE THE CO-ORDS ARE AT THE ARROWS BASE.
\adjx=\arrowlength      \adjy=\arrowlength
\multiply\adjx by \XDIR \multiply\adjy by \YDIR  % Set in \SETDIR
\slanttest(\adjx,\adjy)
\global\advance\boxlengthx by \adjx    \global\advance\boxlengthy by \adjy
\fi
\ifnum\phantomswitch=0\put(\boxlengthx,\boxlengthy){\vector(\XDIR,\YDIR){0}}\fi
}  % END OF \drawarrow.
\gdef\SETFRONTSTEM{
\EITHERSTEM=\FRONTSTEM   \advance\EITHERSTEM by \BACKSTEM
\ifdim\EITHERSTEM>0pt
\global\stemlengthx=\stemlength   \global\stemlengthy=\stemlength
\global\absstemlength=\stemlength
\SETDIR
\gslanttest(\stemlengthx,\stemlengthy)
\gslanttest(\absstemlength,\REG)  % the \REG is to use up the parameter space.
\ifnum\XDIR=0 \stemlengthx=0 \fi
\ifnum\YDIR=0 \stemlengthy=0 \fi
\global\multiply\stemlengthx by \XDIR
\global\multiply\stemlengthy by \YDIR
\ifdim\FRONTSTEM=1pt
\ifnum\phantomswitch=0
          \put(\pfrontx,\pfronty){\line(\XDIR,\YDIR){\absstemlength}}\fi
\global\advance\plengthx by \stemlengthx
\global\advance\plengthy by \stemlengthy
\global\advance\PFRONTx by \stemlengthx
\global\advance\PFRONTy by \stemlengthy
\global\advance\pmidx by \stemlengthx
\global\advance\pmidy by \stemlengthy
\global\advance\pbackx by \stemlengthx
\global\advance\pbacky by \stemlengthy
\ifnum\LTYPE=3
\global\photonfrontx=\PFRONTx  \global\photonfronty=\PFRONTy
\global\photonbackx=\pbackx    \global\photonbacky=\pbacky
\fi  % END LTYPE
\ifnum\LTYPE=4
\global\gluonfrontx=\PFRONTx  \global\gluonfronty=\PFRONTy
\global\gluonbackx=\pbackx    \global\gluonbacky=\pbacky
\fi  % END LTYPE
\fi  % END FRONTSTEM
\fi  % END EITHERSTEM
}    % end of \SETFRONTSTEM
\gdef\SETBACKSTEM{
\ifdim\BACKSTEM=1pt
\ifnum\phantomswitch=0
       \put(\pbackx,\pbacky){\line(\XDIR,\YDIR){\absstemlength}}\fi
\global\advance\plengthx by \stemlengthx
\global\advance\plengthy by \stemlengthy
\global\advance\pbackx by \stemlengthx
\global\advance\pbacky by \stemlengthy
\fi  % END BACKSTEM
\global\stemlength=275  \FRONTSTEM=0pt  \BACKSTEM=0pt % Reset default switches.
}    % END OF \SETBACKSTEM
\gdef\drawloop#1[#2#3](#4,#5){  %RECURSIVE.
\input LOOPS  % contains loops definitions
\drawloop#1[#2#3](#4,#5)}
\Feynmanlength  % Set length scale to centipoints.

\newcount\scalarlength
\newcount\scalarlengthx
\newcount\scalarlengthy
\newcount\scalarfrontx  %}(x,y) co-ord of left of scalar
\newcount\scalarfronty  %}
\newcount\scalarbackx
\newcount\scalarbacky
\gdef\ALLscalar{
\global\scalarfrontx=\particlefrontx   % READ IN FROM FEYNMAN \selectscalar
\global\scalarfronty=\particlefronty   % READ IN FROM FEYNMAN \selectscalar
\numlineparts = 1      \numupperunits=\unitboxnumber
\ifcase\LINECONFIGURATION
\global\upperlineadjx=-200     \global\upperlineadjy=0
\slanttest(\seglength,\gaplength)   %SEE FEYNMAN22.TEX.
\gdef\upperunitbox{\line(\XDIR,\YDIR){\seglength}}
\else \UNIMPERROR % etc.
\fi
\global\unitboxwidth=\seglength  \global\advance\unitboxwidth by \gaplength
\global\multiply \unitboxwidth by \XDIR
\global\unitboxheight=\seglength  \global\advance\unitboxheight by \gaplength
\global\multiply \unitboxheight by \YDIR
\global\particleadjustx=\gaplength \global\multiply\particleadjustx by \XDIR
\global\particleadjusty=\gaplength \global\multiply\particleadjusty by \YDIR
\negate\particleadjustx   \negate\particleadjusty   % SUBTRACT from linelength
\setparticle  %SCALAR8
\global\scalarlengthx=\particlelengthx  %SCALAR8
\global\scalarlengthy=\particlelengthy  %SCALAR8
\ifnum\boxlengthx > 50000
\message{   *** WARNING *** Scalar of length in excess of 50000cp
requested!}\fi
\ifnum\boxlengthy > 50000
\message{   *** WARNING *** Scalar of length in excess of 50000cp
requested!}\fi
\global\scalarbackx=\pbackx      \global\scalarbacky=\pbacky   %SCALAR8
}

\title{$O(4)$ Expansion of the Ladder Bethe-Salpeter Equation}
\author{Taco Nieuwenhuis and J. A. Tjon}
\address{Institute for Theoretical Physics, University of Utrecht,
Princetonplein 5,\\
P.O. Box 80.006, 3508 TA Utrecht, the Netherlands.}
\begin{document}

\draft
\baselineskip=1.5\baselineskip

\maketitle
\begin{center}
\today
\end{center}

\begin{abstract}
The Bethe-Salpeter amplitude is expanded on a hyperspherical basis, thereby
reducing the original 4-dimensional integral equation into an infinite
set of coupled 1-dimensional ones. It is shown that this representation
offers a highly accurate method to determine numerically  the bound state
solutions.
For generic cases only a few hyperspherical waves are needed to achieve
convergence, both for the ground state as well as for radially or orbitally
excited states.
The wave function is reconstructed for several cases and in particular
it is shown that it becomes independent of the relative time in
the nonrelativistic regime.
\end{abstract}

\section{Introduction}
\label{intro}

Over the past 40 years the Bethe-Salpeter equation has been
the traditional starting point for relativistic field theoretical studies of
two-body bound states\cite{bands,nak,itzu}. In view of the numerical complexity,
most studies are
restricted to the ladder approximation. Within this approximation the 
two-body amplitude obeys a 4-dimensional integral equation that can easily be 
transformed into a Fredholm-type eigenvalue problem by a Wick
rotation\cite{wick}.
Except for the Wick-Cutkosky model \cite{wick,cut} where two scalar particles
interact 
through the exchange of a massless particle, very limited analytically
obtained information is known about the
solutions of the ladder Bethe-Salpeter equation.
Moreover, despite its fairly transparent structure, it has proven to be
a nontrivial numerical problem to find accurate solutions of it
for a more general type of interaction. 

Many studies have been carried out to determine the Bethe-Salpeter
eigenvalues and vertex functions numerically. Most efforts start
with the reduction of the 4-dimensional integral equation into a
2-dimensional one by introducing a partial wave decomposition,
thereby exploiting rotational symmetry\cite{levine,zuil}. The 
2-dimensional integral equation is then solved by standard discretization
methods. Although the resulting matrix equations can be solved numerically,
one has often to deal with very large matrices, especially when the spin
complication of the external particles has to be accounted for.
In addition, this happens also when the exchanged mass becomes small. To
account properly for the long range character of the interaction, a larger
number of integration points is needed in that case.
Variational calculations \cite{zwart,vosko} have also been done and proven to be
a
powerful technique. They are, however, harder to apply to studies of excited
states.

In this work we consider an alternative method for solving the
ladder Bethe-Salpeter equation based on a hyperspherical expansion\cite{fabre}.
Instead of doing a standard partial
wave decomposition, thereby expanding the amplitude on a
basis of $O(3)$ spherical harmonics, we expand the amplitude
on an $O(4)$ basis. As a result the ladder Bethe-Salpeter equation
is transformed into an infinite set of coupled 1-dimensional
integral equations. This representation clearly has the advantage that the
integral
equations are substantially easier and more accurately solvable, provided that
the number of channels is limited. The idea behind this method is in fact quite
old 
and some approximated, lowest order results have already been published a long
time ago
\cite{gourdin}. Here we will systematically improve upon the lowest order
calculations and solve the full set of integral equations until the desired
accuracy has been achieved. For typical calculations
only very few hyperspherical waves are needed in practice in order
to find a reasonable accuracy. In particular when the
exchanged mass or the binding energy is sizable 
as compared to the constituent masses,
only 1 or 2 hyperspherical waves are required to reach
convergence. 

In Sect.\ \ref{thebse} we discuss the ladder Bethe-Salpeter equation 
and introduce the hyperspherical expansion
to derive the resulting coupled set of 1-dimensional integral equations.
In Sect.\ \ref{numres} we present the numerical results. The rate of convergence
of the
eigenvalues and vertex functions is studied and shown to be fast which
renders the method an efficient one. 
Finally, in Sect.\ \ref{concl} we show that the method can also
be used to reconstruct the solutions in the case of (2+1) dimensions and some
concluding remarks are made. 
Two appendices contain the explicit expressions of the kernel of the
Bethe-Salpeter equation on the hyperspherical basis, both for (3+1) as
well as for (2+1) dimensions.

\section{The Ladder Bethe-Salpeter Equation}
\label{thebse}

For definiteness, let us consider the bound state problem of two scalar
particles with mass $m$ in $(d-1)$ spatial and 1 time dimensions. 
In a relativistic field theory 
a two-particle bound state with mass $M$ corresponds to a solution of the
homogeneous  Bethe-Salpeter equation  (BSE) at total momentum $P$ with
$P^2=M^2$. After a Wick rotation \cite{wick}, the BSE for the corresponding
vertex function $\Phi (q)$  has the general structure
\begin{equation}
\Phi (q) = \frac{1}{(2\pi )^d}\int {\rm d}^d q'\: V(q,q') S\left(q'
\right)\Phi (q'), 
\label{bse}
\end{equation}
where $q$ is the relative momentum, $V(q,q')$ is the set of all 2-particle
irreducible 
diagrams and $S(q)$ is the free 2-particle Green function. In the CM-frame ($P
=({\bf 0},\sqrt{s})$)
we have for $S(q)$
\begin{eqnarray}
S(q) & = & \frac{1}{(q^2+m^2-\mbox{$\frac{1}{4}s$})^2+(q\cdot P)^2},\nonumber\\
& = & \frac{1}{\left(q^2 + m^2 - \mbox{$\frac{1}{4}$}s
\right)^2+q^2s\cos^2\chi_{Pq}},
\label{sscalar}
\end{eqnarray}
with $\chi_{Pq}$ being the angle between the (Euclidean) vectors $P$ and $q$ in
$d$-dimensional momentum space. For a given $V(q,q')$, Eq.\ (\ref{bse}) only 
supports solutions for values of $\sqrt{s} < 2 m$ that correspond to bound
states.

\begin{figure}[t]
\begin{picture}(30000,6000)(-9000,0)
\global\Yone=15000 \global\Xone=300
\THICKLINES

\drawline\fermion[\E\REG](0,1000)[4000]
\drawline\fermion[\E\REG](0,5000)[4000]
\global\gaplength=365
\global\seglength=365
\drawline\scalar[\S\REG](2000,5000)[6]

\put(5700,2700){\makebox(0,0)[bl]{+}}

\drawline\fermion[\E\REG](8000,1000)[6000]
\drawline\fermion[\E\REG](8000,5000)[6000]
\global\gaplength=365
\global\seglength=365
\drawline\scalar[\S\REG](10000,5000)[6]
\global\gaplength=365
\global\seglength=365
\drawline\scalar[\S\REG](12000,5000)[6]

\put(15700,2700){\makebox(0,0)[bl]{+}}

\drawline\fermion[\E\REG](18000,1000)[8000]
\drawline\fermion[\E\REG](18000,5000)[8000]
\global\gaplength=365
\global\seglength=365
\drawline\scalar[\S\REG](20000,5000)[6]
\global\gaplength=365
\global\seglength=365
\drawline\scalar[\S\REG](22000,5000)[6]
\global\gaplength=365
\global\seglength=365
\drawline\scalar[\S\REG](24000,5000)[6]

\put(27700,2700){\makebox(0,0)[bl]{+\hspace{0.5cm}
$\cdots\cdots$}}
%\put(12000,-1000){\makebox(0,0)[bl]{Ladder series}}

\end{picture}
\caption{General structure of the ladder diagrams.}
\label{fig1}
\end{figure}
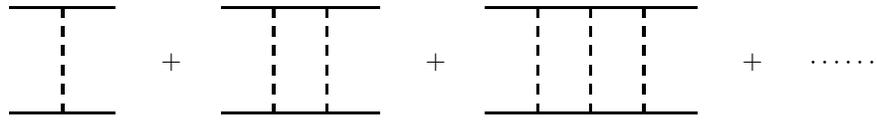

In this paper we will confine ourselves to the particular example of
two scalar
particles with mass $m$, interacting through the exchange of another scalar
particle with mass
$\mu$. In the ladder approximation we have
\begin{eqnarray}
V(q,q') & = & g^2\frac{1}{(q-q')^2+\mu^2},\nonumber \\
& = & g^2 \frac{1}{q^2+q'^2+\mu^2-2|q||q'|\cos\gamma_{qq'}},
\label{vscalar}
\end{eqnarray}
with $\gamma_{qq'}$ the angle between the relative momenta $q$ and $q'$.
Eq.\ (\ref{bse})
sums up the set of ladder diagrams depicted in Fig.\ \ref{fig1}.
In this work we focus our attention to the (3+1)-dimensional case. 
Appendix \ref{spherbse3} contains the essential expressions for the
(2+1)-dimensional case and the generalization to other space-time dimensions
can be made analogously.

Usually Eqs.\ (\ref{bse}-\ref{vscalar}) are solved by noting that for $d=4$ 
they exhibit $O(3)$-symmetry and
as a result it is profitable to expand the amplitudes on a basis of $O(3)$
spherical 
harmonics $Y_{lm}(\theta,\phi )$. This
reduces the ladder BSE to a 2-dimensional integral equation in the 3-momentum
${\bf | q |}$
and the relative energy variable $q_4$. Since for 
$s=0$ the BSE has an additional O(4)-symmetry, which as a
consequence reduces it to a single 1-dimensional integral equation in that
particular
case, it is tempting to explore the convergence rate of an expansion in terms
of hypersphericals in 4 dimensions. Such a method should in
particular work very well for the case of strong binding. Starting from
Eq.\ (\ref{bse}) 
we may expand the vertex function $\Phi(q)$ in appropriate 4-dimensional
spherical harmonics. 
From Eq.\ (\ref{sscalar}) it is clear that
for $s \neq 0$ the BSE-kernel is not $O(4)$ invariant. Hence
the expansion in hyperspherical harmonics will not lead to a complete separation
of the generalized angular variables. 

In (3+1) dimensions the suitable set of spherical harmonics
is given by:
\begin{equation}
Z_{klm}(\chi,\theta,\phi) = X_{kl}(\chi)
Y_{lm}(\theta,\phi),
\label{Yklmdef}
\end{equation}
with
\begin{eqnarray}
X_{kl}(\chi) & \equiv & \sqrt{ \frac{2^{l+1}}{\pi}
\frac{(k+1)(k-l)!l!^2}{(k+l+1)!}}\: \sin^l\chi\:
C_{k-l}^{l+1}(\cos\chi),
\label{Ckldef}\\
Y_{lm}(\theta, \phi)& \equiv & (-1)^m\sqrt{\frac{2l+1}{4\pi}
\frac{(l-m)!}{(l+m)!}}\:P_l^m(\cos\theta)\:{\rm e}^{{\rm i} m\phi}.
\label{Ylmdef}
\end{eqnarray}
Here $C_{k-l}^{l+1}(x)$ are the Gegenbauer polynomials of degree $k-l$ and
order $l+1$ and $P_l^m(x)$ are the associated Legendre functions of
the first kind of degree $l$ 
and order $m$ \cite{abro}. For a fixed value of $k$, the index $l$ assumes
integer values from 0 to $k$, while $m$ goes from $-l$ to $l$.
Standard properties of these functions allow one to
prove completeness and orthogonality for the set $Z_{klm}(\chi,\theta,\phi)$
with respect to the scalar product at hand
\begin{equation}
\int_0^{2\pi}\!\!\!\!{\rm d}\phi \int_0^\pi\!\!{\rm d}\theta\sin\theta\int_0^\pi
\!\!{\rm d}\chi\sin^2\chi Z_{klm}(\chi,\theta,\phi)Z^*_{k'l'm'}
(\chi,\theta,\phi)=\delta_{kk'}\delta_{ll'}\delta_{mm'}.
\label{Zortho}
\end{equation}
By expanding $\Phi(q)$ and $V(q,q')$ in Eq.\ (\ref{bse}) as
\begin{eqnarray}
\Phi (q) & = & \sum_{klm} \Phi_{klm}(|q|)Z_{klm}(\chi,\theta,\phi)
\label{phiklm}\\
V(q,q') & = & \sum_k V_k(|q|,|q'|)C_k^1(\cos\gamma_{qq'})
\label{Vklm}
\end{eqnarray}
and using the addition theorem for the 4-dimensional
hyperspherical harmonics
\begin{equation}
C_k^1(\cos\gamma) = \frac{2\pi^2}{k+1}\sum_{lm} Z_{klm}(\chi,\theta,\phi)
Z^*_{klm}(\chi',\theta',\phi'),
\label{Zaddit}
\end{equation}
where $\gamma$ is the angle between the two unit vectors $(\chi,\theta,\phi)$ and
$(\chi',\theta',\phi')$, we obtain after some straightforward
algebra an infinite set of coupled 
1-dimensional integral equations:
\begin{equation}
\Phi_{klm}(q) = \frac{1}{8\pi^2(k+1)}\sum_{k'}\int_0^\infty\!\!\!\!
{\rm d}q' q'^3\:
V_{k}(q,q')\:S_{kk'}^{l}(q')\:\Phi_{k'lm}(q').
\label{bse3+1klm}
\end{equation}
Here the matrix $S_{kk'}^{l}(q)$ is given by
\begin{equation}
S_{kk'}^{l}(q) =\int_0^\pi \!\!{\rm d}\chi\sin^2\chi\:
X_{kl}(\chi)\:X_{k'l}(\chi)\:S(|q|,\cos\chi).
\label{Skkdef}
\end{equation}
For our particular case the matrix-elements $V_k (q,q')$ and $S_{kk'}^{l}(q)$ can
explicitly be computed and are given in Appendix \ref{spherbse4}.

Eq.\ (\ref{bse3+1klm}) exhibits several symmetries that allow
for a physical interpretation and which lead to simplifications
in the numerical treatment.
Since $V_{k}(q,q')$ and $S_{kk'}^l(q)$ are independent of the magnetic
quantum number $m$, the solutions
of (\ref{bse3+1klm}) do not depend on it. Hence there is
a $(2l+1)$-fold degeneracy present which is a manifestation of the
cylindrical symmetry. Furthermore, Eq.\ (\ref{bse3+1klm}) is diagonal in
the orbital quantum number
$l$ which implies the conservation of spatial angular momentum, in contrast to
the nonconservation of the generalized angular momentum 
connected to the quantum number $k$. Considering the behavior 
under parity transformations $x\rightarrow -x$ of the Gegenbauer
polynomials:
$C_{k-l}^{l+1}(-x)= (-1)^{k-l}C_{k-l}^{l+1}(x)$ leads to the
conclusion that $S_{kk'}^l(q)$ is only nonzero if $k+k'$ is
even. As a result only states of equal parity are coupled to each other
in (\ref{bse3+1klm}) and consequently parity is conserved. Finally, one can show
that
$S_{kk'}^l(q)$ vanishes unless $k\geq l$ and $k'\geq l$. Hence, for a
fixed value of $l$, the quantum numbers of the generalized angular
momentum assume values $k = l + 2 j$ and $ k'=l+2j'$, with $j \in
0,1,2,\ldots$.

\section{Numerical Results}
\label{numres}

Having transformed the original 4-dimensional integral equation (\ref{bse})
into an infinite set of coupled 1-dimensional ones,
our aim is now to find those values of $s<4 m^2$ for which Eq.\ (\ref{bse3+1klm})
allows solutions. As noted these would correspond to bound states with mass
$M=\sqrt{s}$. In particular it is interesting to investigate how many
hyperspherical waves are required in order to achieve a certain accuracy and to
see how this number depends on the
properties of the particular state one is considering.

\subsection{Numerical Implementation}
\label{implem}

In order to evaluate the integral in
Eq.\ (\ref{bse3+1klm}), we project the infinite interval $q=[0,\infty)$ on to the
finite
unit interval $z=[0,1]$ through the transformation: $q\rightarrow z/(1-z)$.
The integral can then simply be computed by Gauss-Legendre quadratures
\cite{numrep}. 
\begin{equation}
\int_{0}^1\!\!{\rm d}x f(x)= \lim_{N\rightarrow\infty}
\sum_{i=1}^N w_if\left(x_i\right),
\label{gaussjacobi}
\end{equation}
with weights $w_i$ and abscissas $x_i$.
Next we truncate the infinite sum over $k'$ in Eq.\ (\ref{bse3+1klm}) at a finite
value of $k'=k_{\rm max}$. As a result we have replaced the kernel of
the integral equation
by an approximated finite matrix ${\bf K}$ of dimension $(N\cdot k_{\rm max})
\times (N\cdot k_{\rm max})$ of which we seek the eigenvalues. 
After having computed $\det ({\openone} - {\bf K})$ numerically, we apply
Newton's method \cite{numrep} in order to find
the roots of $\det ({\openone} - {\bf K})$ as a function of $\sqrt{s}$ for a
fixed value
of $g$ or {vice versa}. Typical values of $N=60$ have been used in the actual
calculations, except for the case of smallest $\mu$, where $N=200$ was needed
to get the same relative accuracy as the other ones. Finally we checked for a 
large number of states that we obtained identical results as zur Linden
\cite{linden}.

\subsection{Convergence of the Eigenvalues}
\label{converge_spec}

Since the BSE is only $O(4)$-invariant for $\sqrt{s}=0$,
the rate of convergence of the hyperspherical method is expected to show a
significant
dependence on the value of the bound state mass $\sqrt{s}\equiv M$. We study the
convergence rate by keeping
$M$ fixed at different values 
and determining the value of the dimensionless coupling constant
$\lambda = g^2/4\pi m^2$ that yields
a bound state of these masses. 

In Table \ref{table_mu} we present a comparison of calculations for different
values
of $M$ and $\mu$. For the calculations we consider the ground state where
the radial quantum number $n=1$ and the system is in an $s$-state ($l=0$).
We indeed find for a given $\mu$ that a larger value of $k_{\rm max}$ is
required for increasing $M$ in order to obtain a certain accuracy.
Except for the smallest value of $\mu/m=0.001$ and the weak binding
case of $M/m=1.999$, it is remarkable to see how fast the method converges. 
The latter exceptional case corresponds to the nonrelativistic situation
where the binding is small as compared to the rest masses of the
constituents.
A reasonable relative accuracy of $10^{-4}$ can rather easily be obtained
by considering only a few hyperspherical waves. The convergence is extremely good
for moderate
values of $\mu/m\simeq 1$ where only two hyperspherical waves suffice for the
desired
accuracy.
Comparing the predictions for the case of $\mu/m=0.001$ with
those of the Wick-Cutkosky model we find that they agree within
4 digits, except for the weak binding situation $M/m=1.999$. The 
poorer correspondence for the latter case is understandable since the
exchange mass, which vanishes in the Wick-Cutkosky case, is of the order of the
binding
energy of the system here. Hence its influence on the long range structure
of the bound state is expected to be significant, while it is negligible for
the other situations.

Next we wish to study how the above pattern depends on the orbital quantum number
$l$. For this
investigation we keep $\mu$ fixed at $\mu/m=0.1$ and we vary $l$ at different
values of
the bound state mass $M$. The results of this calculation is presented in
Table \ref{table_l}. From this table we see, for each value of $M$,
that the rate of convergence is virtually independent of the orbital quantum
number $l$.

Finally, solutions of radially excited states have been studied. Some results
are shown in Table \ref{table_n}.
Also in this case we find that the hyperspherical method works very well.
The convergence rate is found to be only very weakly dependent on the
radial quantum number $n$.

 From the above investigations, we may conclude that for the computation of the
spectrum, the rate of convergence of the hyperspherical method is, as
expected, dependent on the bound state mass $M$ and on the value
of the exchanged mass $\mu$. 
The convergence rate is in general very fast except for small
values of the exchanged mass $\mu$. This is not surprising since
again we expect that the long range character of the force for $\mu
\rightarrow 0$ should be reflected in the rate of convergence.
In this connection, it should be noted that keeping a finite
number of the hyperspherical waves, the resulting kernel of the
BSE is compact even in the limit $\mu \rightarrow 0$.
For performing the integrals in an accurate way, more
integration points are needed when $\mu$ becomes small.
This may be attributed to the appearance of a cusp in
$V_k(q,q')$ at $q=q'$ when $\mu$ is identically 0.

\subsection{Hyperspherical Components of the Wave Functions}
\label{converge_wave}

Clearly, the BSE (\ref{bse}) with kernel $\bf K$ has formally the structure of a
generalized eigenvalue problem
\begin{equation}
\lambda \Phi = {\bf K} \Phi.
\label{eigen}
\end{equation}
After discretization of the integral in the BSE we may find for a given energy 
$\sqrt{s}$ the eigenvalues and eigenvectors of the matrix $\bf K$ using standard
procedures.
The physical solutions can simply be obtained by varying $s$ such that we get
an  eigenvalue $\lambda=1$. 
Once the eigenvectors have been determined of Eq.
(\ref{eigen}) we immediately obtain the wave function $\Psi (q)$ through the
relation
\begin{equation} 
\Psi (q) = S (q)  \Phi (q) 
\end{equation}

In the actual reconstruction of the ground state vertex function we have used the
so-called power method \cite{mal}, which is considerably faster than
the procedures for obtaining general eigenvectors. It simply consists of the
iteration procedure
\begin{equation}
\Phi^{(n+1)} = {\bf K} \Phi^{(n)}.
\label{iter}
\end{equation}
One can readily show, that the limiting function $\Phi^{(n\rightarrow\infty )}$ 
becomes proportional to the eigenvector with largest 
eigenvalue $\lambda$. This eigenvalue is simply given by 
$\lambda=\lim_{n\rightarrow\infty}\Phi^{(n+1)}/\Phi^{(n)}$.
To accelerate the iterative procedure we have chosen a starting vector
$\Phi^{(0)}$ which incorporates the asymptotic behavior of the 
different hyperspherical waves
\begin{eqnarray}
\Phi_{k0}(q) & \stackrel{q\rightarrow 0}{\sim} & q^k\label{rmPhi},\\
& \stackrel{q\rightarrow \infty}{\sim} & q^{-k-2},\label{rmPhiq8}
\end{eqnarray}
More specifically, we choose $\Phi_{k0}^{(0)}(q)$ according to:
\begin{equation}
\Phi_{k0}^{(0)}(q) = \frac{q^k}{\left(1+q^2\right)^{k+1}}.
\label{rmPhi0}
\end{equation}
The method was used as an additional check on all numerically obtained
eigenvalues.

Obviously the eigenvectors can be obtained in this way up to a normalization.
We have normalized the wave functions according to the well known BSE
normalization condition given by \cite{nak}
\begin{equation}
\int\!\!\frac{{\rm d}^4q}{(2\pi)^4}\frac{{\rm d}^4q'}{(2\pi)^4}\Psi (q)
\frac{\partial}{\partial P_\mu } \left[ (2\pi)^4\delta^4 (q-q') S^{-1}(q) + 
V(q,q')\right]\Psi^*(q') {=} 2P_\mu
\label{norm}
\end{equation}
Inserting the expansion (\ref{phiklm}) in (\ref{norm}) we find in the
CM-frame for each partial wave $l$:
\begin{equation}
\frac{1}{2(2\pi)^4}\sum_{kk'}\int_0^\infty\!\!\!\!{\rm d}q
\:q^3\Psi_{kl}(q)\left[
2q^2A_{kk'}^l
-\delta_{kk'}\left( q^2+m^2-\mbox{$\frac{1}{4}$}s\right)
\right]\Psi_{k'l}(q) {=}1
\label{normklm}
\end{equation}
where we have defined:
\begin{eqnarray}
\Psi_{kl}(q)&=&\sum_{k'}S_{kk'}^l(q)\Phi_{k'l}(q)\label{psik}\\
A_{kk'}^l& = & \int_0^\pi\!\!{\rm d}\chi\, \sin^2\chi \: X_{kl}(\chi)\:
X_{k'l}(\chi)\,\cos^2\chi\label{offdia}
\end{eqnarray}
In Appendix \ref{spherbse4} the explicit expression for $A_{kk'}^l$ is given. 

\begin{figure}[t]
\epsfxsize=16cm
\epsfysize=10.5cm
\epsffile{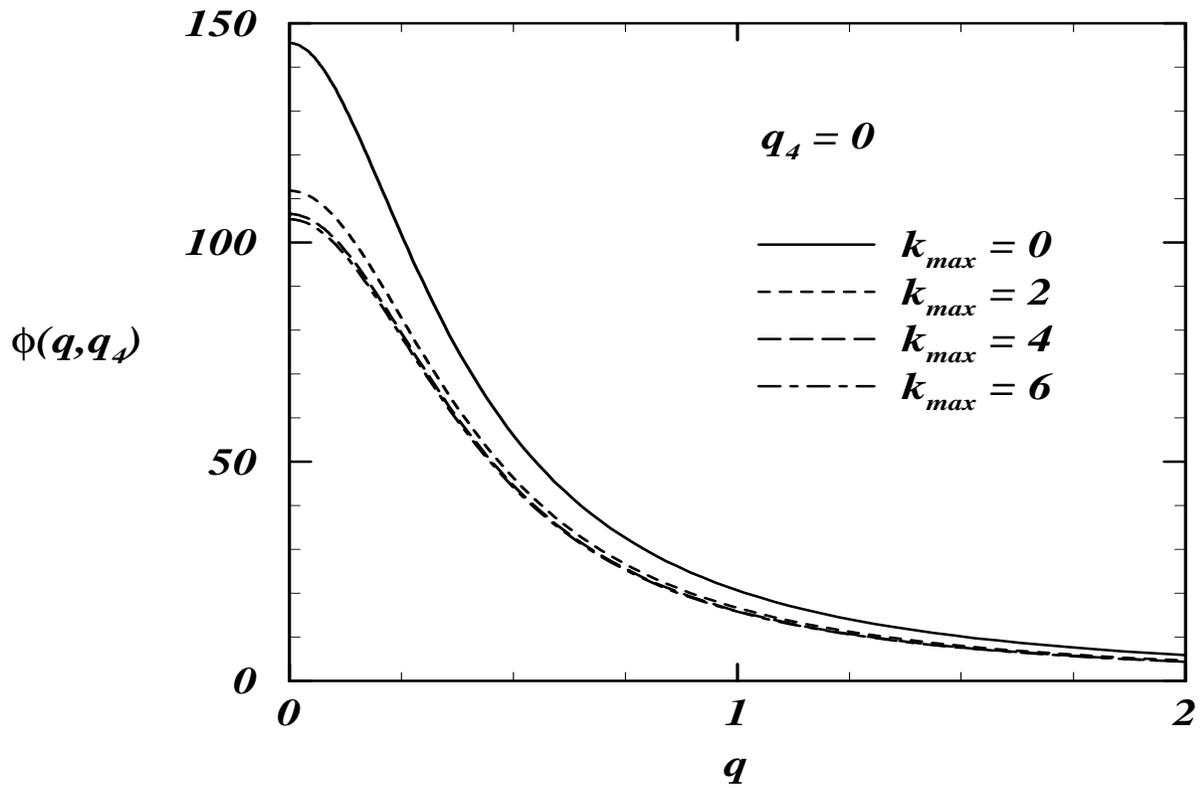}
\vspace{1cm}
\epsfxsize=16cm
\epsfysize=10.5cm
\epsffile{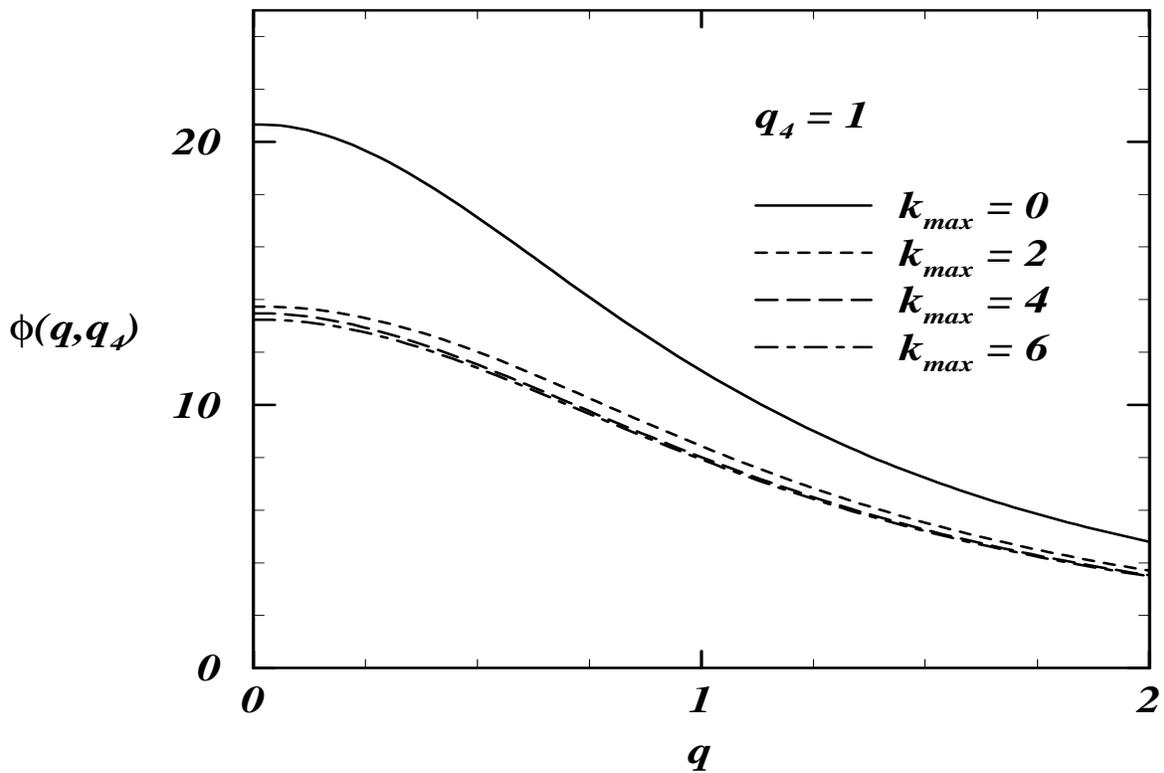}
\caption{Rate of convergence of hyperspherical expansion of the vertex functions
$\Phi ({\bf q},q_4)$ for
$\mu/m=0.1$, $M/m=1.9$ and $l=0$ at $q_4 =0$ and 1. Note the different scales on
the vertical axis.}
\label{figvert}
\end{figure}

\begin{figure}[t]
\epsfxsize=16cm
\epsfysize=10.5cm
\epsffile{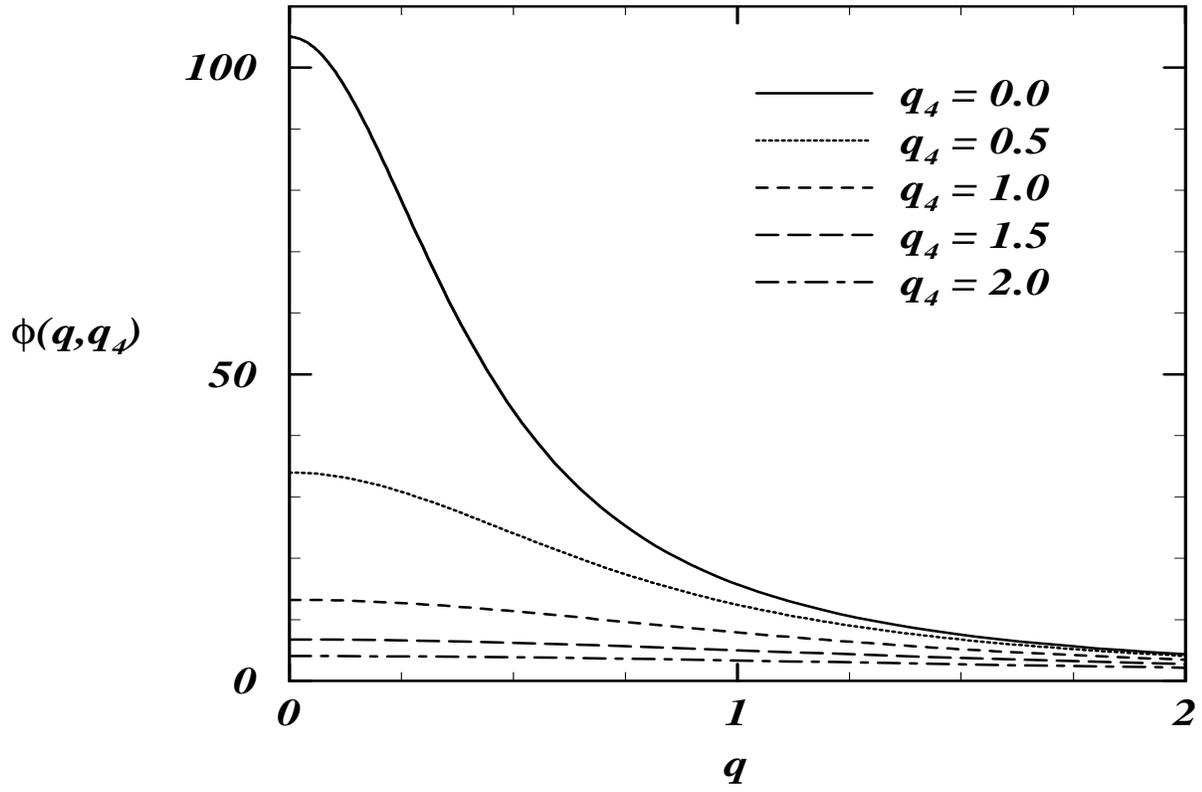}
\caption{The $|{\bf q}|$ and $q_4$ dependence 
of the  vertex function $\Phi (q,q_4)$ for $\mu/m=0.1$, $M/m=1.9$ and
$l=0$ at various values of the relative energy $q_4$.}
\label{figdiffo}
\end{figure}

In Fig.\ \ref{figvert} we show as typical examples the rate of convergence of
the hyperspherical method for the vertex function $\Phi (|{\bf q}|,q_4)$ of the
ground state
for two values of the relative energy $q_4$ in the case of
a bound state mass $M=1.9m$. For the whole $|{\bf q}|-q_4$ region 
we need only 4 hyperspherical waves for this case to get a reasonably converged 
result for the vertex function. As we can see from Fig.\ \ref{figdiffo}, the
$|{\bf q}|$- and $q_4$-dependence of the calculated $\Phi$ are smooth and the
range of
fall-off in these two variables are similar.

\begin{figure}[t]
\epsfxsize=16cm
\epsfysize=10.5cm
\epsffile{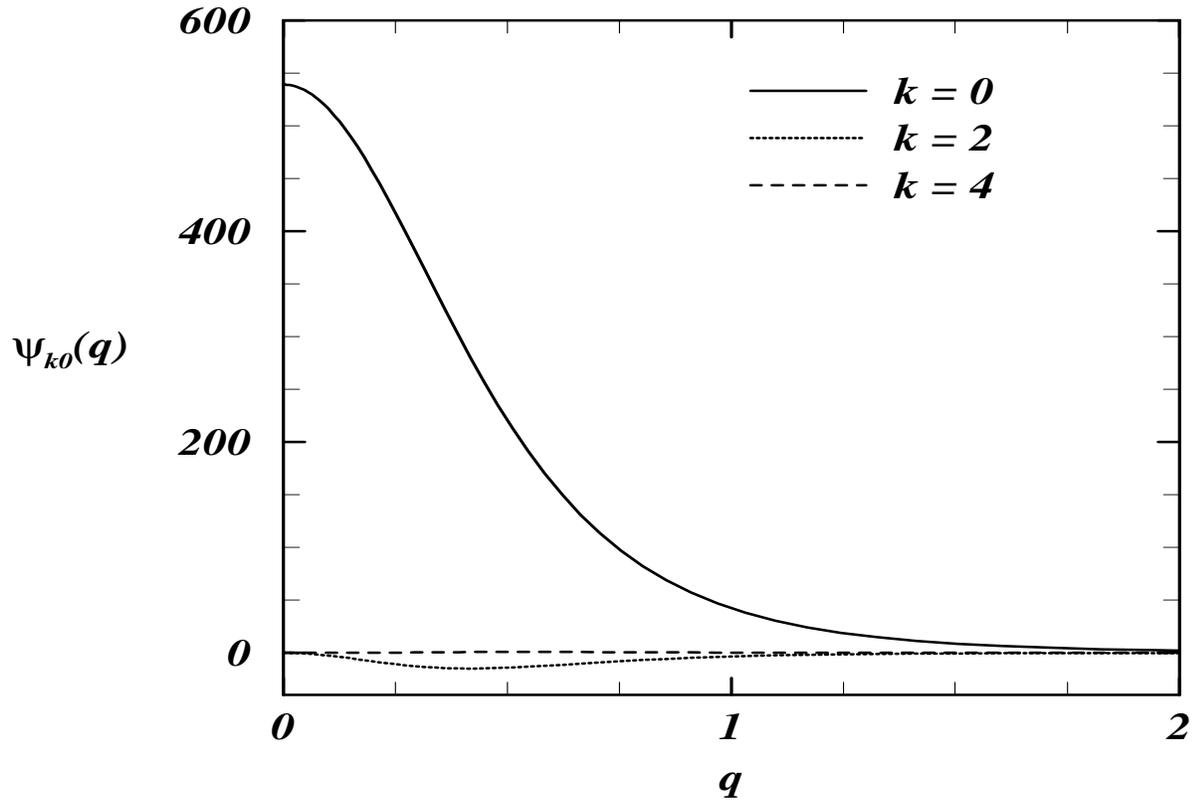}
\vspace{1cm}
\epsfxsize=16cm
\epsfysize=10.5cm
\epsffile{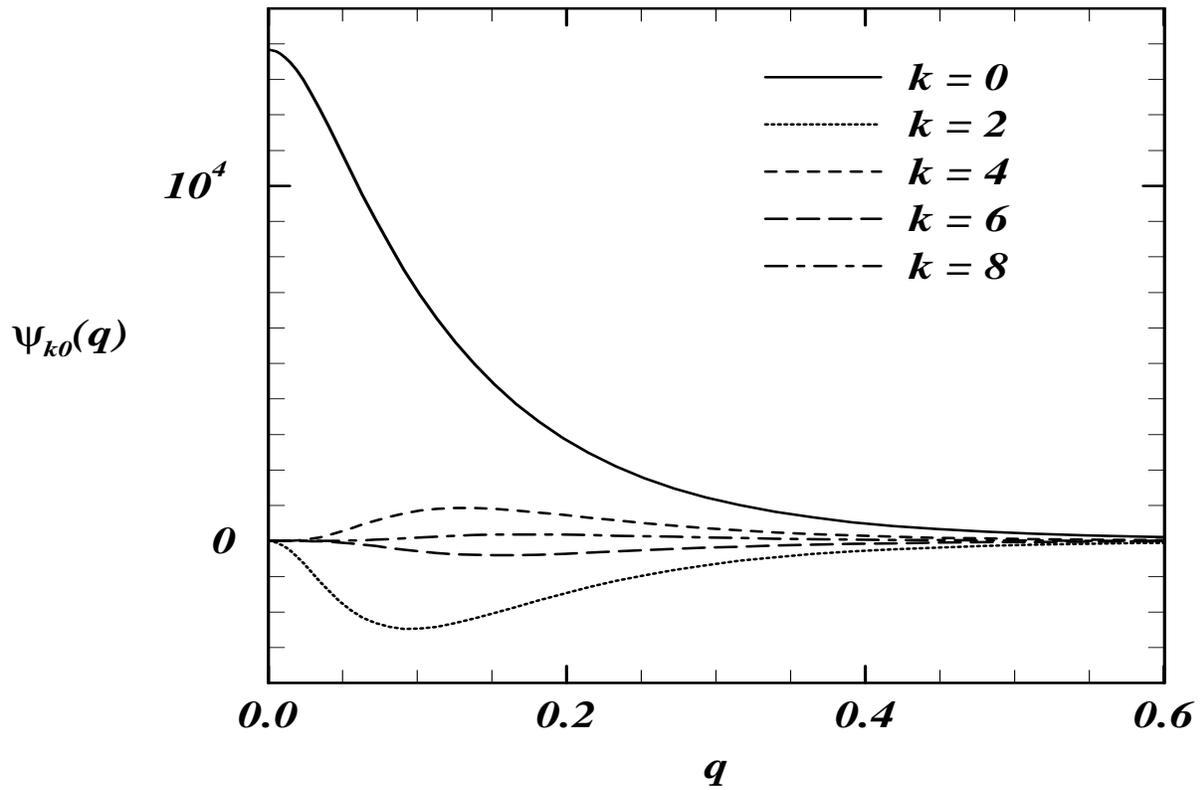}
\caption{Hyperspherical components of the wave function
$\Psi_{k0}(q)$ for 
$\mu/m=0.1$ and $l=0$. The
upper graph is for the case  $M/m=1$ while the
lower one contains the solutions for $M/m=1.9$.}
\label{fig4}
\end{figure}

The rapid convergence of the hyperspherical method is related to the magnitude
of the hyperspherical components in the wave function.
In Fig.\ \ref{fig4} we display for $l=0$ and $\mu/m=0.1$ the components
of the ground state wave function for the cases $M/m=1$ and $1.9$. The
corresponding
coupling constants are given by $\lambda=20.678$ respectively $\lambda=5.227$. 
For the latter case we see that 
the contributions of higher ($k\geq 2$) hyperspherical waves are indeed
more significant than for the former situation, in agreement with the slower rate
of convergence of the eigenvalue. Moreover, from the figure we see that the
range of the fall-off in $q$ is very different for the two cases. This is
predominantly caused by the $s$-dependence of the free 2-particle Green
function and related to the size of the composite system. The hyperspherical
components of the corresponding vertex functions exhibit a more comparable
range and the non-leading ones are relatively more suppressed in magnitude.

\begin{figure}[t]
\epsfxsize=16cm
\epsfysize=10.5cm
\epsffile{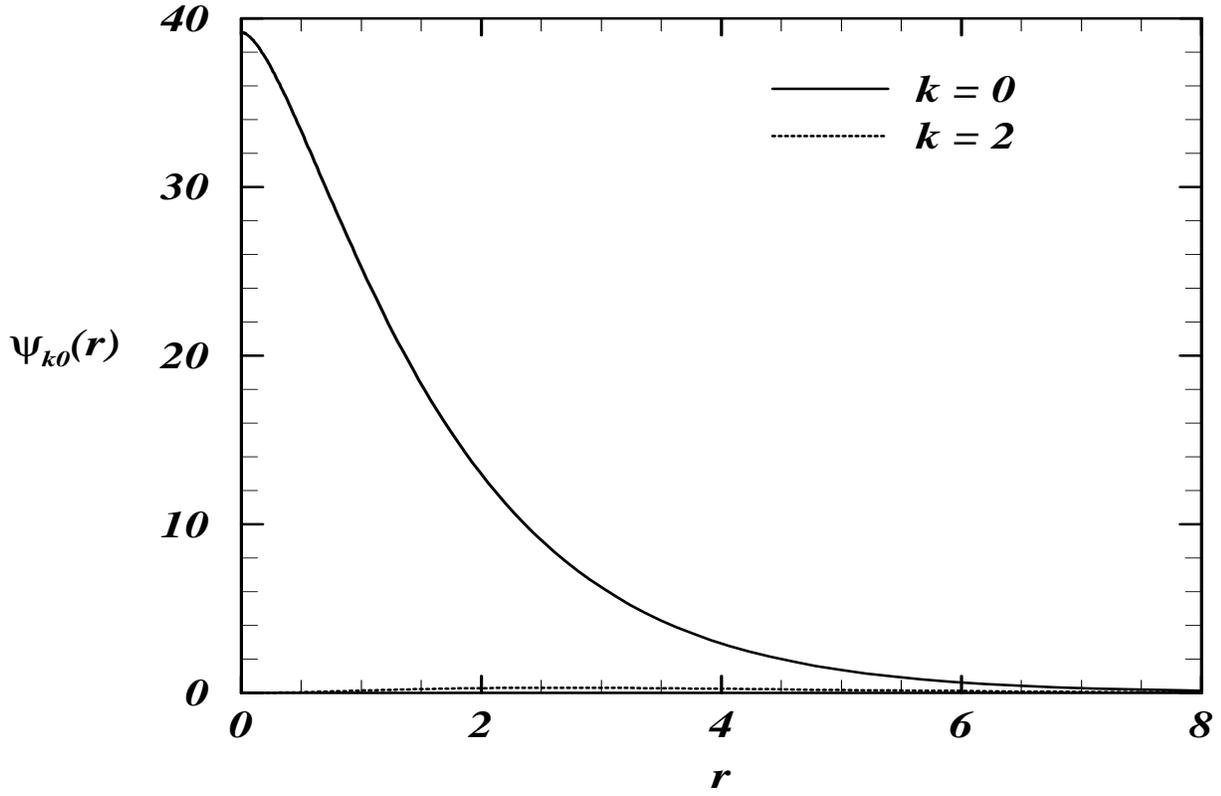}
\vspace{1cm}
\epsfxsize=16cm
\epsfysize=10.5cm
\epsffile{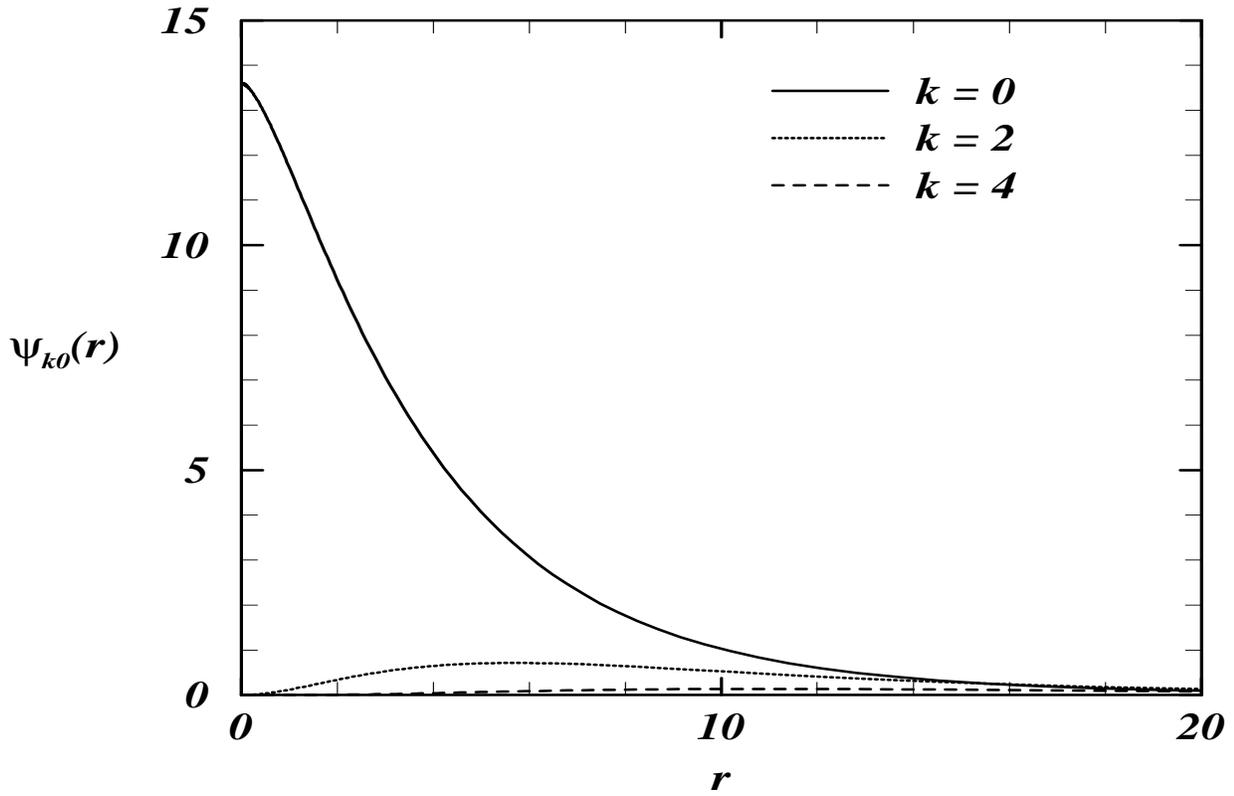}
\caption{ Hyperspherical components of the  configuration space wave function
for $\mu/m=0.1$ and $l=0$. The
upper graph shows the components ${\hat \Psi}_{k0}(r)$ for $M/m=1$ while the
lower one contains the solutions for $M/m=1.9$.}
\label{fig5}
\end{figure}

\begin{figure}[t]
\epsfxsize=16cm
\epsfysize=10.5cm
\epsffile{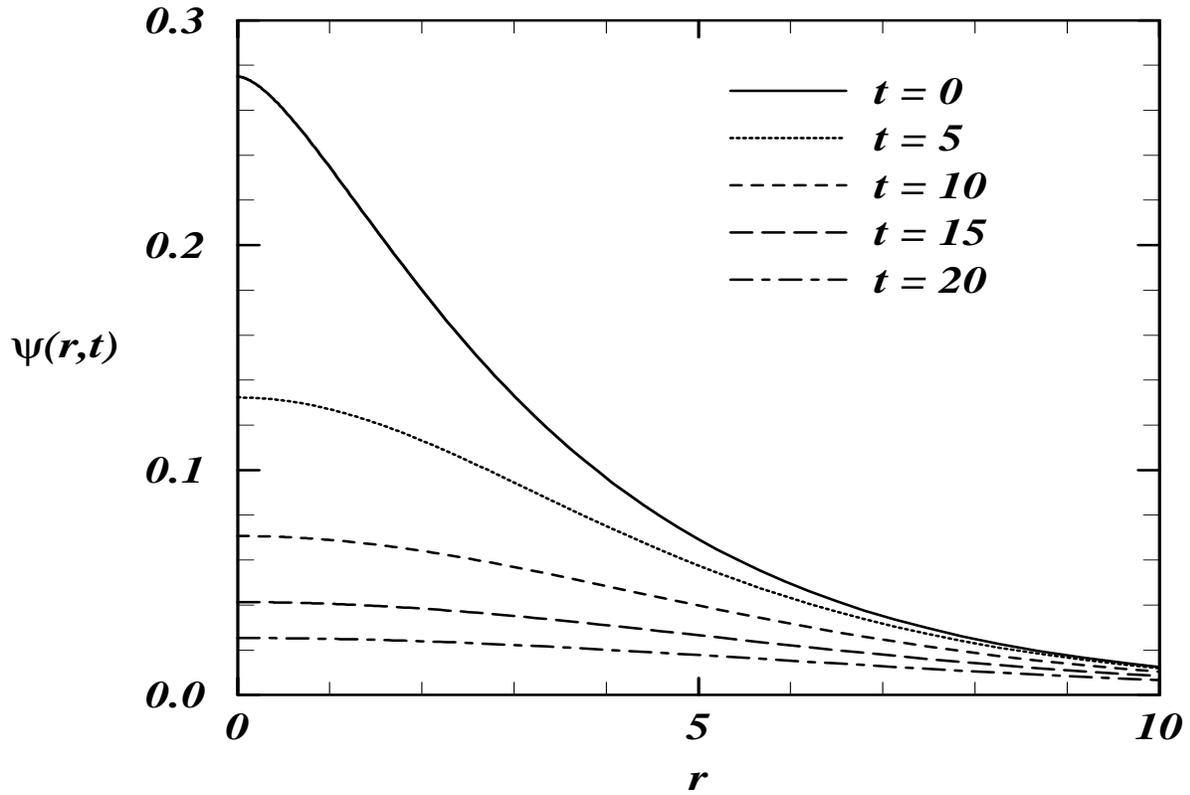}
\caption{The $|{\bf r}|$ and $t$ dependence 
of the  configuration space wave function $\Psi_0 (|{\bf r}|,t)$ for
$\mu/m=0.1$, $M/m=1.9$ and $l=0$ at various values of the relative time $t$.}
\label{figdifft}
\end{figure}

In the weak binding regime we expect both particles to be nearly on their
mass-shell, 
which results for the wave function $\Psi $ in a strongly peaked structure in the
$q_4$-variable as compared to the typical fall-off in the $|{\bf q}|$-variable. 
As a consequence we expect that the wave function in the configuration space
will exhibit an independence on the relative time $t$ in the
nonrelativistic limit. We have studied this transition region.
The wave functions in configuration space can be obtained 
from the above ones via Fourier transformation.
With the help of the representation of the complex exponential
\begin{equation}
{\rm e}^{{\rm i} q\cdot r} = 4\pi^2 \sum_{klm} \frac{J_{k+1}(qr)}{ qr}
Z_{klm}(\chi_q,\theta_q,\phi_q)Z^*(\chi_r, \theta_r,\phi_r),
\label{eiqr}
\end{equation}
we find that for a fixed $l$, the configuration space wave functions
are given by
\begin{equation}
\Psi_l(|{\bf r}| ,t) = \frac{1}{4\pi^2\sqrt{{\bf r}^2+t^2}}
\sum_k {\rm i} ^k X_{kl}\!\!
\left({\rm arccos}\frac{t}{\sqrt{{\bf r}^2+t^2}}\right) 
\int_0^\infty \!\!\!\!{\rm d}q
\:\: q^2
J_{k+1}\left(q\sqrt{{\bf r}^2+t^2}\right) \Psi_{kl}(q)
\label{phil}
\end{equation}
The integral Eq.\ (\ref{phil}) was performed by Gauss-Legendre integration.
In order to compute the wave functions $\Psi_l (|{\bf r}|,0)$ at relative time
zero,
one needs the value of $X_{kl}(\frac{1}{2}\pi)$
\begin{equation}
C_{k-l}^{l+1}(0) = (-1)^{\frac{1}{2}(k-l)}\left(\begin{array}{c}
\frac{1}{2}(k+l) \\
\frac{1}{2}(k-l)\end{array}\right)\hspace{0.5cm}
\Longrightarrow \hspace{0.5cm} X_{kl}(\mbox{$\frac{1}{2}$}\pi)  = 
(-1)^{\frac{1}{2}(k-l)}
\sqrt{ \frac{2^{l+1}}{\pi}
\frac{(k+1)(k-l)!(\frac{1}{2}(k+l))!}{(k+l+1)! (\frac{1}{2}(k-l))!}}
\label{cklhpi}
\end{equation}
whereas the determination of the  wave function $\Psi_l (0,t)$
at zero relative distance requires $X_{kl} (0)$
\begin{equation}
C_{k-l}^{l+1}(1)= \delta_{l0}(k+1) \hspace{1cm}\Longrightarrow \hspace{1cm}
X_{kl}(0) = \delta_{l0}\sqrt{\frac{2}{\pi}}(k+1),
\label{ckl0}
\end{equation}
We display in Fig.\ \ref{fig5} the computed configuration space components
\begin{equation}
{\hat \Psi}_{kl}(r) =
\frac{1}{4\pi^2 r}\int_0^\infty \!\!\!\!{\rm d}q \:\: q^2
J_{k+1}\left(qr\right) \Psi_{kl}(q) \label{phil1}
\end{equation}
of the ground state, again for
$M/m=$1.0 and 1.9  with $\mu/m=0.1$. Comparing this with Fig.\ \ref{fig4}
we see that the pronounced structure of the higher components is
relatively more suppressed and its
strength is spread out to larger values of $\bf r$. The asymmetric behaviour of
the configuration space wave function $\Psi_l$ from Eq.\ \ref{phil} as a function
of $|{\bf r}|$ and $t$ can clearly be seen in Fig.\ \ref{figdifft} for the
case of $M/m$=1.9, $l$=0.
Its fall-off in $t$ is twice as slow as compared to that in $|{\bf r}|$.

\begin{figure}[t]
\epsfxsize=13cm
\epsfysize=8.5cm
\epsffile{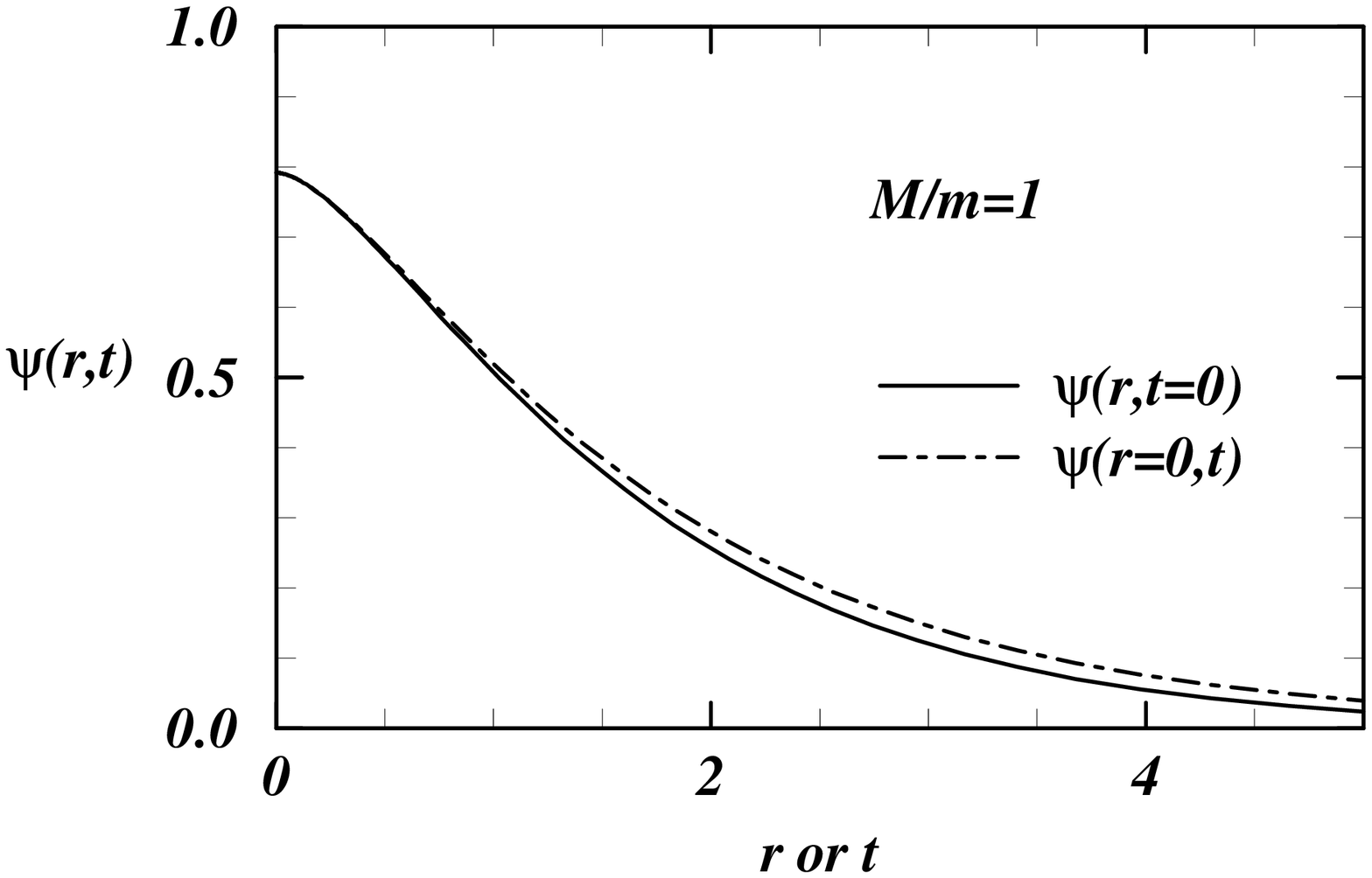}
\vspace{-2cm}
\epsfxsize=13cm
\epsfysize=8.5cm
\epsffile{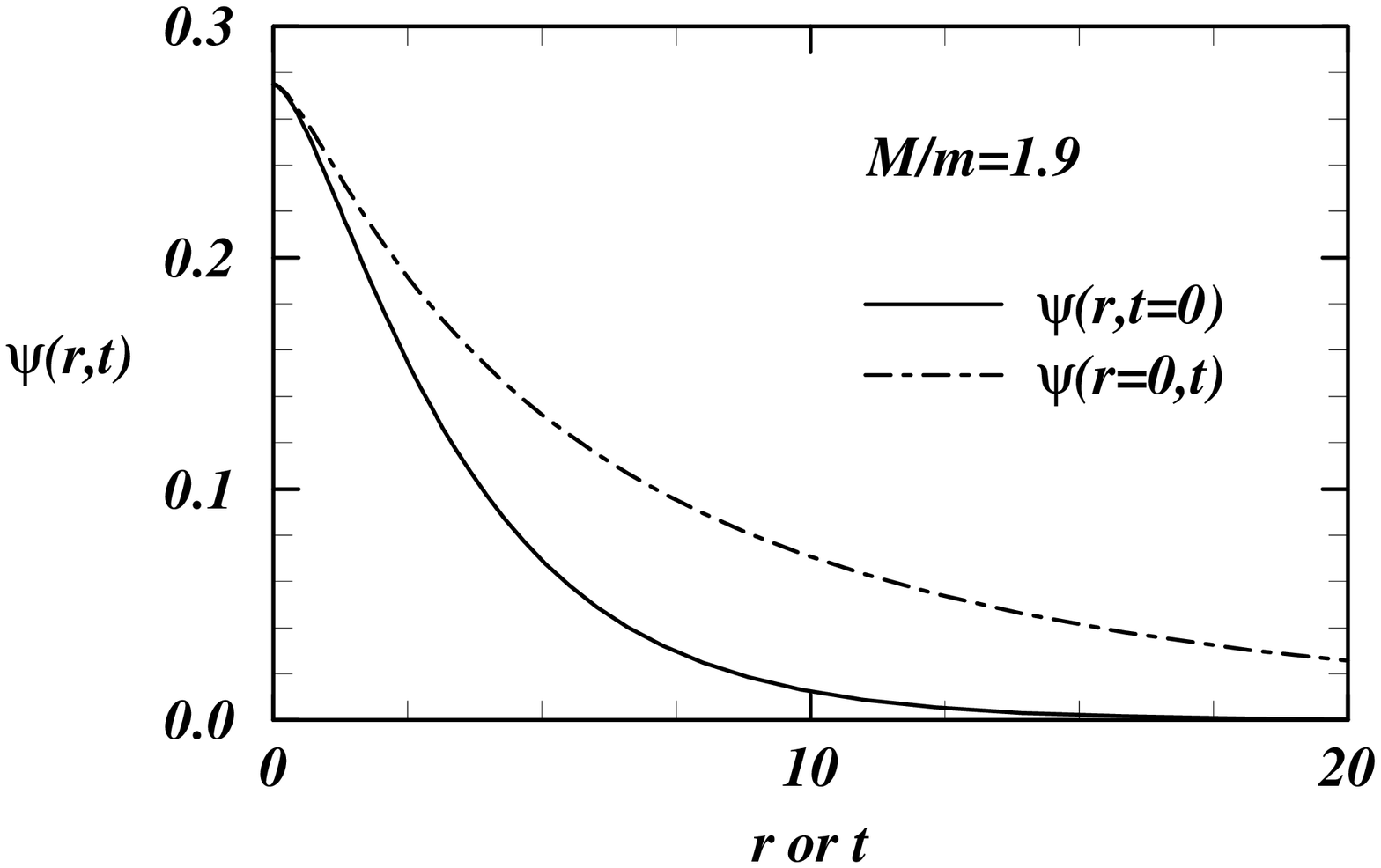}
\vspace{-2cm}
\epsfxsize=13cm
\epsfysize=8.5cm
\epsffile{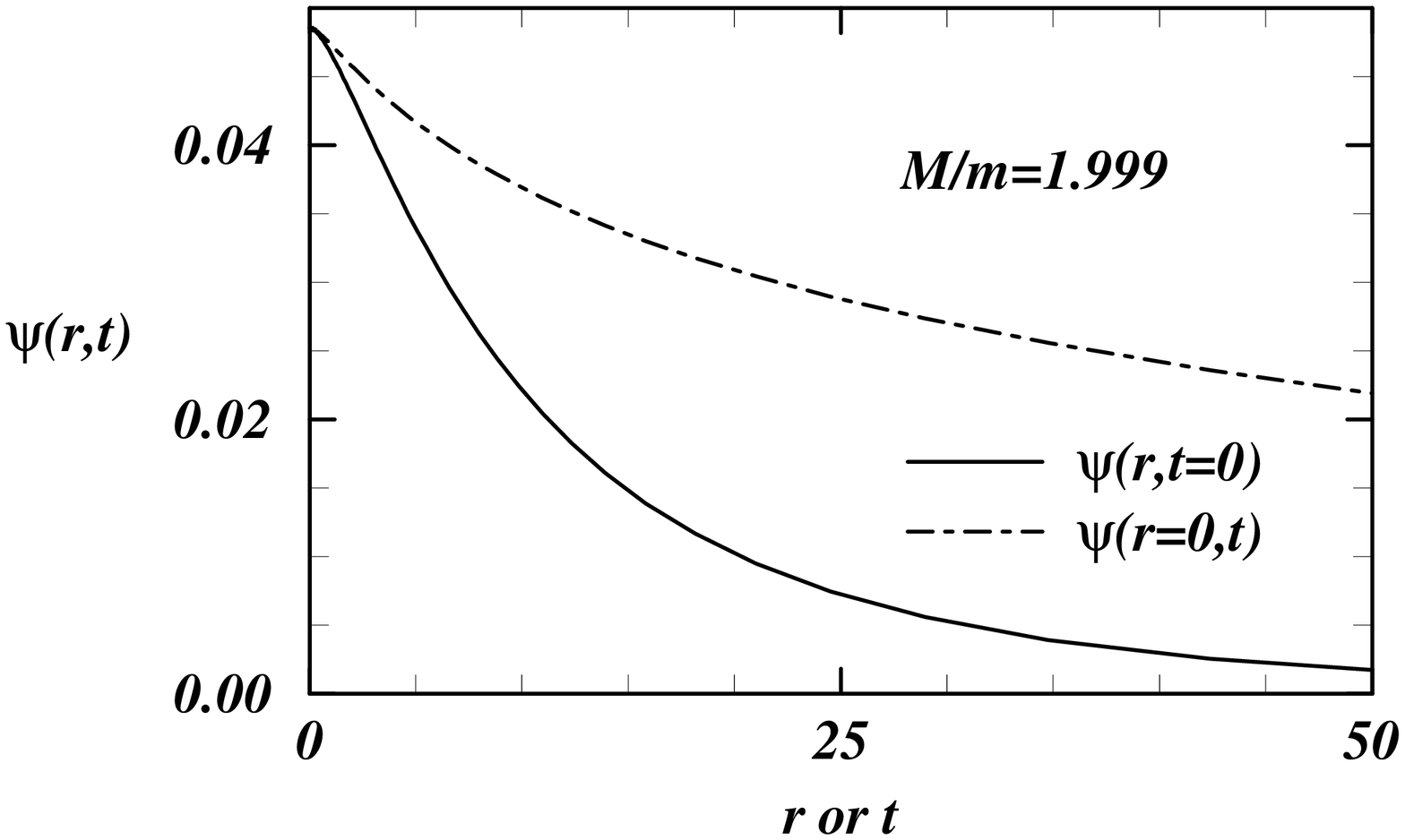}
\caption{Comparison of the relative time zero wave function $\Psi_0(|{\bf
r}|,t=0)$ with the
relative distance zero wave function $\Psi_0(r=0,t)$ for $\mu/m=0.1$ and $l=0$.
In the nonrelativistic limit $M/m\rightarrow 2$, $\Psi_0(|{\bf r}|,t)$ is seen to
become less dependent on $t$.}
\label{fig7}
\end{figure}

Finally, it is of some interest to study the role of the relative time in the
transition region to weak binding.
In Fig.\ \ref{fig7} we show three graphs with $M/m=1$, 1.9 and 1.999 where we
compare the dependence on the Euclidean relative time $t$ to that of the relative
distance $|{\bf r}|$.
We expect that in going to the region where the binding energy becomes
negligible as compared to the total mass of the system, to recover the
Schr\"odinger wave functions and to lose the dependence on the relative time
which is
inherently absent in a nonrelativistic description. 
For $M/m\rightarrow 2$ we indeed observe that the dependence of $\Psi_0 (|{\bf
r}|,t)$ on
$t$ becomes substantially weaker as compared to that on $\bf r$, i.e., the wave
functions clearly show the expected independence on the relative time $t$.

\section{Concluding Remarks}
\label{concl}

We have investigated the hyperspherical expansion for solving the ladder
Bethe-Salpeter
equation in (3+1) dimensions. It was found that it constitutes a powerful and
efficient method for
obtaining an accurate determination of the eigenvalues as well as the
eigenfunctions. This is true both for the ground state as well as for
excited states.
We stress that all calculations presented here took at most
one minute of CPU-time on a HP735-workstation.
Except for the limiting cases of vanishing exchange mass or small
binding energy, only a few hyperspherical waves suffice for computing the
solutions
to the ladder Bethe-Salpeter equation within an accuracy that is sufficient for
most
practical purposes. Furthermore, it was found that the eigenfunctions could be
reconstructed in a straightforward fashion.

The method can equally well be generalized to the case of
$d$-dimension. In particular, we may consider the
case of 2 spatial and 1 time dimensions. The explicit form of the
kernel of the BSE for this case is given in Appendix \ref{spherbse3}.
In Table \ref{table_ir} we show the converged results for the
dimensionless coupling constant $\lambda = g^2/4 \pi
m$ for fixed values of the bound state mass and $\mu\rightarrow 0$.
Similar rates of convergence properties are found as for the (3+1)
dimensional case, i.e., for weak binding and smaller 
exchange mass $\mu$ more terms in the hyperspherical expansion are 
needed. From Table \ref{table_ir} we see that
the limit of $\mu\rightarrow0$ is smooth and no logarithmic confinement is
found, as would have been expected from a nonrelativistic analysis.
Such a logarithmic behavior can readily be seen using
the frequently adopted procedure to
obtain a static nonrelativistic potential $V_{\rm NR}(|{\bf r}|)$
from a relativistic interaction:
\begin{equation}
V_{\rm NR}(|{\bf r}|)  \equiv  \int_{-\infty}^\infty \!\!\!\! {\rm d}t
\:V(|{\bf r}|,t).\label{vnr}
\end{equation}
For the Yukawa interaction in (2+1) dimensions we find consequently:
\begin{equation}
V_{\rm NR, Yukawa}(|{\bf r}|) = -\frac{g^2}{2\pi}K_0(\mu|{\bf r}|).
\label{yuk}
\end{equation}
The asymptotic behavior of the Bessel
function for small arguments $K_0(z)\sim -\log (\frac{1}{2}z)$, suggests
clearly that the nonrelativistic Coulomb interaction in (2+1) dimensions is
logarithmically confining. 
According to table \ref{table_ir} we find however the limiting
behavior of $\mu \rightarrow 0$ to be smooth for the relativistic
ladder BSE. Hence no confinement is seen in that limit. 
This paradoxical situation was investigated
in \cite{tacomu} where it is found that a nonuniformity exists 
between the nonrelativistic
limit $m\rightarrow\infty$ and the Coulomb limit $\mu\rightarrow 0$.

Given the very encouraging results we found in this study of scalar particles, we
expect the hyperspherical expansion to be a powerful and accurate 
method to solve more realistic physical models such the ones that include spin
degrees of
freedom. In particular, we do expect that it can equally well be applied
with good convergence properties for cases like the spinor-spinor interaction.
Moreover, it should be interesting to investigate the applicability of the
method to the scattering case.

\appendix

\section{Hyperspherical Projection of the BSE$_{3+1}$}
\label{spherbse4}

In this appendix we give the explicit formulae for the kernel of the
(3+1) dimensional BSE that occur in the hyperspherical expansion.
Most expressions are obtained with the help of standard
properties of special functions, as can be found, e.g., in
Ref.\ \cite{abro,prud,grad}

The projection of the interaction on the hyperspherical basis
can be performed analytically
\begin{eqnarray}
V_k (q,q') & = & g^2\frac{2}{\pi}\int_0^\pi\!\!{\rm d}\chi\sin^2\chi
\:C_k^1(\cos\chi)\:V(q,q',\cos\chi)
\nonumber \\
& = & g^2 \frac{4}{\left( \Lambda_++\Lambda_-\right)^2}\left( 
\frac{\Lambda_+-\Lambda_-}{\Lambda_++\Lambda_-}\right)^k
\label{vk}
\end{eqnarray}
where $\Lambda_\pm \equiv \sqrt{(q\pm q')^2+\mu^2}$. 

Substituting Eq.\ (\ref{Ckldef}) in Eq.\ (\ref{Skkdef}) and introducing
$x=\cos \gamma_{Pq}$, we may rewrite $S_{kk'}^l(q)$ as
\begin{equation}
S_{kk'}^l(q) =  \frac{2^{l+1}l!^2}{\pi}\sqrt{\frac{(k+1)(k-l)!}{(k+l+1)!}}
\sqrt{\frac{(k'+1)(k'-l)!}{(k'+l+1)!}}
\int_{-1}^1\!\!\!{\rm d}x \left(1-x^2\right)^{l+1/2}
C_{k-l}^{l+1}(x)C_{k'-l}^{l+1}(x) S(q,x).
\label{Skkela}
\end{equation}
After some tedious algebra this reduces to
\begin{equation}
S_{kk'}^l(q) = 
\frac{2\sqrt{2}{\rm i} l!}{z\sqrt{\pi}}
\sqrt{\frac{(k+1)(k-l)!}{(k+l+1)!}} \sqrt{\frac{(k'+1)(k'-l)!}{(k'+l+1)!}}
{\rm e}^{-(\frac{1}{2}l-\frac{1}{4})\pi{\rm i}}
(1+z^2)^{\frac{1}{2}l+\frac{1}{4}}C_{k_+}^{l+1}({\rm i}
z)Q_{k_-+l+\frac{1}{2}}^{l+\frac{1}{2}}
({\rm i} z)
\label{Skkelaex}
\end{equation}
with 
\begin{equation}
z\equiv \frac{q^2+m^2-\frac{1}{4}s}{q\sqrt{s}}
\label{zdef}
\end{equation}
and $k_+\equiv {\rm max}(k,k')$ and
$k_-\equiv {\rm min}(k,k')$. Despite the appearance of the imaginary unit `${\rm
i}$'
in Eq.\ (\ref{Skkelaex}), $S_{kk'}^l(q)$ is always real.
For $l=0$ a more transparent form can be found from
(\ref{Skkelaex}).
\begin{equation}
S_{kk'}^0 (q) = (-1)^{(k_+-k_-)/2}\frac{1}{sq^2z\sqrt{1+z^2}}
\left[\left(\sqrt{1+z^2}-z\right)^{k_++k_-+2}
+(-1)^{k_-}
\left(\sqrt{1+z^2}+z\right)^{k+--k_-}\right]
\label{skk0}
\end{equation}

The matrix-element $A_{kk'}^l$, defined in Eq.\ (\ref{offdia}) as the expectation
value of the operator $\cos^2\chi$ between two spherical harmonics, can be
computed with the help of the recurrence relation for Gegenbauer polynomials
\begin{equation}
2(n+m+1)xC_{n+1}^m(x) = (n+2m)C_n^m(x)+(n+2)C_{n+2}^m(x),
\label{reccu}
\end{equation}
which yields the following expression for $A_{kk'}^l$
\begin{eqnarray}
A_{kk'}^l & =&  \mbox{$\frac{1}{4}$} \left[\delta_{k,k'}\left(
\frac{(k-l)(k+l+1)}{
k(k+1)}+\frac{(k-l+1)(k+l+2)}{(k+1)(k+2)}\right)\right.\nonumber\\
&&\hspace{4cm}\left.+\delta_{k,k'+2}\frac{(k-l)(k+l+1)(k'-l+1)(k'+l+2)}{k(k+1)(k'+1)(k'+2)}
+k\leftrightarrow k' \right]
\label{akkl}
\end{eqnarray}

\section{Hyperspherical Projection of the BSE$_{2+1}$}
\label{spherbse3}

In (2+1) dimensions the spherical harmonics
are the $Y_{lm}(\theta, \phi)$ functions. The projection of the ladder BSE in
this case closely follows that of the (3+1) dimensional case. After expanding
the vertex function and the interaction on the hyperspherical basis
\begin{eqnarray}
\Phi (q)& =& \sum_{lm} \Phi_{lm}(q)Y_{lm}(\theta,\phi),\label{phiexp}\\
V(q,q') & = & \sum_{l}V_l(|q|,|q'|)P_l(\cos\gamma_{qq'}),\label{vexp}
\end{eqnarray}
and applying the appropriate addition theorem 
\begin{equation}
P_l (\cos \gamma ) = \frac{4\pi}{2l+1}\sum_m Y_{lm}(\theta,\phi)Y^*_{lm},
\theta'\phi')
\label{add3}
\end{equation}
we find 
\begin{equation}
\Phi_{lm}(q)=\frac{1}{2\pi^2(2l+1)}\sum_{l'}\int_0^\infty\!\!\!\!
{\rm d}q' q'^2 V_l(q,q') S_{ll'}^m(q')\Phi_{l'm}(q').\label{bselm}
\end{equation}
In Eq.\ (\ref{bselm}) the matrix $S_{ll'}^m(q)$ is defined analogously
to Eq.\ (\ref{Skkdef})
\begin{equation}
S_{ll'}^m(q)=\int_0^\pi\!\!{\rm d}\theta\sin\theta\: P_l^m(\cos\theta)\:
P_{l'}^m(\cos\theta )\: S(|q|,\cos\theta).
\label{Slldef}
\end{equation}
Similar considerations concerning conservation of angular momentum
(with quantum number $m$) and parity as have been made earlier for the (3+1)
dimensional case, apply here as well.

For the one meson exchange diagram we have
\begin{eqnarray}
V_l(q,q') & = & \frac{2l+1}{2}\int_0^\pi\!\!{\rm d}\theta \sin \theta
P_l (\cos\theta) V(q,q',\cos\theta)\nonumber\\
& = & g^2\frac{2l+1}{2qq'}Q_l\left(\frac{q^2+q'^2+\mu^2}{2qq'}\right)
\label{vlexp}
\end{eqnarray}
Moreover, the free 2-particle Green function with $m=0$, $S_{ll'}^0(q)$,
can be evaluated analytically: 
\begin{equation}
S_{ll'}^0(q)=\frac{\sqrt{2l+1}\sqrt{2l'+1}}{4\pi{\rm i}}\frac{1}{sq^2 z}
P_{l_-}({\rm i} z)Q_{l_+}({\rm i} z),
\label{Sllex}
\end{equation}
where $z$ is defined in Eq.\ (\ref{zdef}), while $l_{\pm}$ are
defined according to $l_+={\rm max}(l,l')$ and $l_-={\rm min}(l,l')$.
Unfortunately, we were unable to find an analytical expression
for $S_{ll'}^m(q)$ for arbitrary $m$.
For $m\neq 0$, the integral in Eq.\ (\ref{Slldef}) was
performed numerically.

\newpage
\begin{table}[t]
\caption{Convergence of hyperspherical eigenvalues for $\lambda=g^2/4\pi m^2$ as
a function of $\mu/m$
for $n=1$ and $l=0$. The
numerical errors are at most one unit in the last decimal. When no further values
are given for increasing
$k_{\rm max}$, the final value was found not to change anymore within the desired
accuracy when $k_{\rm max}$
was increased. The case of $\mu/m=0$ is the Wick-Cutkosky
prediction.}
\begin{center}
\begin{tabular}{|lc|c|c|c|c|}
$\mu/m$ & $k_{\rm max}$ & $M/m=0$ & $M/m=1$ & $M/m=1.9$ & $M/m=1.999$ \\ 
\hline
0     &   --          &   25.13      &   20.01      &   4.483       &   0.2854   
     \\
 \hline
0.001 &   0           &   25.13      &   20.02      &   4.675       &   0.3484   
     \\
      &   2           &              &   20.01      &   4.503       &   0.3099   
     \\
      &   4           &              &              &   4.486       &   0.2992   
     \\
      &   6           &              &              &   4.483       &   0.2947   
     \\
      &   8           &              &              &               &   0.2924   
     \\
      &   10          &              &              &               &   0.2912   
     \\ 
      &   12          &              &              &               &   0.2904   
     \\ 
      &   14          &              &              &               &   0.2899   
     \\ 
      &   16          &              &              &               &   0.2896   
     \\ 
      &   18          &              &              &               &   0.2894   
     \\ 
      &   20          &              &              &               &   0.2893   
     \\ 
      &   22          &              &              &               &   0.2892   
     \\ 
      &   24          &              &              &               &   0.2891   
     \\ 
      &   26          &              &              &               &   0.2890   
     \\ 
\hline
0.100 &   0           &   25.80      &   20.69      &   5.374       &   1.0733   
    \\
      &   2           &              &   20.68      &   5.237       &   1.0463   
    \\
      &   4           &              &              &   5.228       &   1.0442   
    \\
      &   6           &              &              &   5.227       &   1.0439   
    \\
      &   8           &              &              &               &   1.0439   
    \\
      &   10          &              &              &               &   1.0438   
    \\ 
\hline
1.000 &   0           &   42.96      &   36.94      &  17.28        &   10.272   
    \\
      &   2           &              &              &  17.23        &   10.251   
    \\
\end{tabular}
\label{table_mu}
\end{center}
\end{table}
\newpage
\begin{table}[t]
\caption{Convergence of hyperspherical eigenvalues for $\lambda=g^2/4\pi m^2$ as
a function of the
orbital quantum number $l$ for $n=1$ and $\mu/m=0.1$. The
numerical errors are at most one unit in the last decimal. When no further values
are given for increasing
$k_{\rm max}$, the final value was found not to change anymore within the desired
accuracy when $k_{\rm max}$
was increased.}
\begin{center}
\begin{tabular}{|cc|c|c|c|c|}
$l$ & $k_{\rm max}$ & $M/m=0$ & $M/m=1$ & $M/m=1.9$ & $M/m=1.999$ \\ 
\hline
 0 &   0           &   25.802     &   20.691     &   5.3743       &   1.0733     
    \\
   &   2           &              &   20.678     &   5.2369       &   1.0463     
    \\
   &   4           &              &              &   5.2279       &   1.0442     
    \\
   &   6           &              &              &   5.2271       &   1.0439     
    \\
   &   8           &              &              &   5.2270       &   1.0439     
    \\
   &   10          &              &              &                &   1.0438     
    \\ 
\hline
 1 &   1           &   160.44     &   127.59     &  33.284        &   10.086     
    \\
   &   3           &              &   127.48     &  32.094        &    9.5620    
    \\
   &   5           &              &              &  32.000        &    9.5043    
    \\
   &   7           &              &              &  31.991        &    9.4957    
    \\
   &   9           &              &              &  31.990        &    9.4942    
    \\
   &   11          &              &              &                &    9.4940    
    \\
   &   13          &              &              &                &    9.4939    
    \\
\hline
 2 &   2           &   668.88     &   531.46     &   143.58       &   55.385     
    \\
   &   4           &              &   531.01     &   137.99       &   51.611     
    \\
   &   6           &              &              &   137.51       &   51.123     
    \\
   &   8           &              &              &   137.46       &   51.042     
    \\
   &   10          &              &              &   137.45       &   51.027     
    \\
   &   12          &              &              &                &   51.024     
    \\ 
   &   14          &              &              &                &   51.023     
    \\ 
\hline
 3 &   3           &   2332.0     &   1857.7     &   525.81        &   237.45    
    \\
   &   5           &              &   1856.2     &   505.29        &   219.63    
    \\
   &   7           &              &              &   503.44        &   217.12    
    \\
   &   9           &              &              &   503.25        &   216.67    
    \\
   &   11          &              &              &   503.23        &   216.59    
    \\
   &   13          &              &              &                 &   216.57    
    \\ 
\end{tabular}
\label{table_l}
\end{center}
\end{table}
\newpage
\begin{table}[t]
\caption{Convergence of hyperspherical eigenvalues for $\lambda=g^2/4\pi m^2$ as
a function of 
the radial excitation quantum number $n$ for $l=0$ and $\mu/m=0.1$. The
numerical errors are at most one unit in the last decimal. When no further values
are given for increasing
$k_{\rm max}$, the final value was found not to change anymore within the desired
accuracy when $k_{\rm max}$
was increased.}
\begin{center}
\begin{tabular}{|cc|c|c|c|c|}
$n$ & $k_{\rm max}$ & $M/m=0$ & $M/m=1$ & $M/m=1.9$ & $M/m=1.999$ \\ 
\hline
 1 &   0           &   25.802     &   20.691     &   5.3743       &   1.0733     
    \\
   &   2           &              &   20.678     &   5.2369       &   1.0463     
    \\
   &   4           &              &              &   5.2279       &   1.0442     
    \\
   &   6           &              &              &   5.2271       &   1.0439     
    \\
   &   8           &              &              &   5.2270       &   1.0439     
    \\
   &   10          &              &              &                &   1.0438     
    \\ 
\hline
 2 &   0           &   80.900     &   64.436     &  16.984        &    4.5741    
    \\
   &   2           &              &   64.388     &  16.424        &    4.2927    
    \\
   &   4           &              &   64.387     &  16.387        &    4.2578    
    \\
   &   6           &              &              &  16.384        &    4.2521    
    \\
   &   8           &              &              &  16.383        &    4.2510    
    \\
   &   10          &              &              &                &    4.2508    
    \\
\hline
 3 &   0           &   169.55     &   135.84     &   38.053       &   12.111     
    \\
   &   2           &              &   134.77     &   35.557       &   10.874     
    \\
   &   4           &              &   134.74     &   35.257       &   10.715     
    \\
   &   6           &              &              &   35.221       &   10.690     
    \\
   &   8           &              &              &   35.217       &   10.686     
    \\
   &   10          &              &              &   35.216       &   10.685     
    \\ 
\end{tabular}
\label{table_n}
\end{center}
\end{table}
\newpage
\begin{table}[t]
\caption{Behavior of the eigenvalues of the BSE in (2+1) dimensions as
$\mu\rightarrow 0$. The calculations are for the ground state.}
\begin{center}
\begin{tabular}{|c|c|c|}
$\mu/m$ & $M/m=1.0$ & $M/m=1.9$ \\ 
\hline
0.1000 &  2.869  &  0.3397    \\
0.0300 &  2.477  &  0.2232    \\
0.0100 &  2.366  &  0.1902    \\
0.0030 &  2.328  &  0.1785    \\
0.0010 &  2.314  &  0.1750    \\
0.0003 &  2.308  &  0.1738    \\
0.0001 &  2.305  &  0.1732    \\
\end{tabular}
\label{table_ir}
\end{center}
\end{table}
\newpage
\newpage
\newpage
\newpage
\newpage
\newpage
\newpage
\end{document}